\title{ Natural gas-fired power plants valuation and optimisation
under L\'{e}vy copulas and regime-switching}
\author{
        Nemat Safarov \\
      \normalsize  Imperial College London\\
        \\
        Colin Atkinson\\
       \normalsize Imperial College London \\
        }
        
        \date{\today}

\documentclass{article}
\usepackage{textcomp}
 \usepackage{amsmath,amssymb}
 \usepackage{cleveref}
\usepackage[titletoc]{appendix}
\usepackage{bbm}
\usepackage{bbold}
\usepackage[square]{natbib}
\usepackage[english]{babel}
\usepackage{graphicx}
\usepackage{caption}
\usepackage{subcaption}
\usepackage[nottoc,notlot,notlof]{tocbibind}
\usepackage{placeins}

\DeclareMathOperator*{\argmax}{arg\,max}
\DeclareMathOperator{\sign}{sign}
\DeclareMathOperator{\minmod}{minmod}
\begin{document}
\maketitle

\newenvironment{proof}[1][Proof]{\begin{trivlist}
\item[\hskip \labelsep {\bfseries #1}]}{\end{trivlist}}
\newcommand{\qed}{\nobreak \ifvmode \relax \else
      \ifdim\lastskip<1.5em \hskip-\lastskip
      \hskip1.5em plus0em minus0.5em \fi \nobreak
      \vrule height0.75em width0.5em depth0.25em\fi}
      
\newtheorem{definition}{Definition}[section]
\newtheorem{theorem}[definition]{Theorem}
\newtheorem{proposition}[definition]{Proposition}
\newtheorem{lemma}[definition]{Lemma}
\newtheorem{condition}[definition]{Condition}



\begin{abstract}
 In this work we analyse a stochastic control problem for the valuation of a natural gas power station while taking into account operating characteristics. Both electricity and gas spot price processes exhibit mean-reverting spikes and Markov regime-switches. The L\'evy regime-switching model incorporates the effects of demand-supply fluctuations in energy markets and abrupt economic disruptions or business cycles. We make use of skewed L\'evy copulas to model the dependence risk of electricity and gas jumps. 
 The corresponding HJB equation is the non-linear PIDE which is solved by an explicit finite difference method. The numerical approach gives us both the value of the plant and its optimal operating strategy depending on the gas and electricity prices, current temperature of the boiler and time.  The surfaces of control strategies and contract values are obtained by implementing  the numerical method for a particular example.

\end{abstract}

\section{Introduction}
The liberalisation of electricity and other energy markets, such as PJM in the United State, UKPX in the United Kingdom, Nord Pool in North Europe,  and JEPX in Japan makes it necessary to incorporate highly volatile spot price dynamics into the optimisation problems for power generation. 

During the traditional regulatory regime, the regulators used to set electricity prices based on
cost of service. Investments in energy generating facilities were allowed to earn a
fixed return through electricity tariffs upon approval by the regulators.
 The economic viability of such
investments and value of power plants could be calculated via a discounted cash flow (DCF) method. But it has been shown that the DCF approach doesn't give correct results as it ignores the opportunity costs (see \cite{dixit1994investment}). Meanwhile, the existence of competitive electricity markets necessitates the valuation of power plants by means of financial instruments. This kind of real option approach to electricity generation was proposed by \cite{deng2001exotic}. \cite{deng2003incorporating} further extend this work by including some physical operating constraints. Note that the omission of physical characteristics usually results in over-valuation.

The real options approach to power plant investment and optimisation problems has become an outstanding research area of different scientific groups.  \cite{pindyck1993investments} applies this method to investigate the investment decisions to build a nuclear power plant. \cite{tseng2002short}  propose forward-moving Monte Carlo simulation with backward-moving dynamic programming technique to solve a multistage stochastic model used to evaluate a power plant with unit commitment constraints.  \cite{nasakkala2005flexibility} obtain a method to calculate thresholds for building a base load plant and upgrading it to a peak load plant 
when spark spread covers emission related costs. \cite{takashima2007investment} investigate the investment problem for  a power plant construction under the assumption that electricity prices are determined by the supply function and the equilibrium quantity.

Power generating assets can be classified as 
\begin{itemize}
\item Nuclear
\item Hydroelectric
\item Fossil
\item Renewable
\end{itemize}
In this work we will only consider fossil  power stations. In general, a thermal power plant is a broader concept than fossil-fueled power generator. But we will use them interchangeably in this paper for simplicity purposes.
Fossil power plants convert heat energy to electric power. More precisely, chemical energy in fossil fuels is first used to heat the water in the boiler. The resulting steam then spins the turbine which drives an electrical generator. The reader can refer to \cite{wood2012power} for more details regarding the operation of thermal power plants.  Coal, natural gas and heating oil are the mostly used fuels for power generation. However, natural gas has become a primary fuel source for thermal power generators because of its low $CO_2$ emissions and flexibility during peak hours. 
The main objective in operating a power plant is to maximise the profit by utilising the difference (or spread) between electricity and natural gas prices. So, the valuation of a gas-fired power plant is closely related to spark spread, which is the most important cross-commodity transaction in power markets and is presented by \cite{deng2001exotic} as
\begin{equation} \label{F1}
S(t)=S_e(t)-cS_g(t),
\end{equation}
where $S_e$ and $S_g$ are electricity and gas spot prices, respectively. heat rate is the amount of gas required to generate 1 MWh of electricity which is defined by heat rate $c$. It is a measure of efficiency. More precisely, the higher is the heat rate, the less efficient is the electricity generation. The heat rate has been usually assumed to be constant for convenience purposes (see \cite{deng2001exotic,benth2011dynamic,meyer2014dynamic}).\cite{deng2003incorporating} generalise this approach by introducing an operating heat rate which changes depending on the discrete output levels. 
\cite{thompson2004valuation} consider the case where $c$ is the amount of natural gas that needs to be burned to maximise the total revenue from electricity generation. In this model $c$ depends continuously on time, boiler temperature, gas and electricity prices and is calculated via dynamic optimisation techniques.

\begin{figure}
	\centering
	\begin{subfigure}[b]{0.90\textwidth}
		\includegraphics[width=\textwidth]{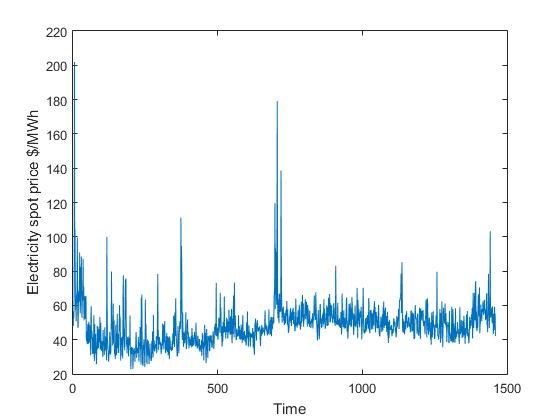}
		\caption{Electricity}
		\label{F1a}
	\end{subfigure}%
	
	~ 
	\begin{subfigure}[b]{0.90\textwidth}
		\includegraphics[width=\textwidth]{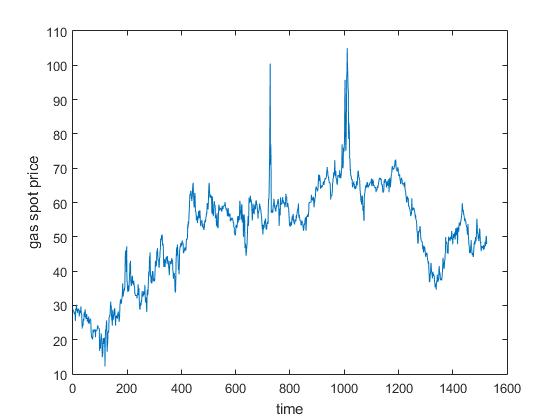}
		\caption{Natural gas}
		\label{F1b}
	\end{subfigure}
	\caption{Day ahead UK spot prices for power and natural gas (NBP) during the time period 4/9/2009 - 4/7/2015 based on APX Power UK and Bloomberg} \label{Fig1}
\end{figure}

The evaluation of a power plant cannot deliver satisfactory results without realistic stochastic models for underlying spot prices (see Fig \ref{Fig1}). The detailed discussion of the related literature was presented in \cite{safarov2016natural}. For completeness purposes we will briefly repeat them here.

A mean-reverting process is a common model for the energy spot price. For example, \cite{schwartz1997stochastic} presented three mean reverting models for commodity spot prices while assuming that stochastic convenience yield and interest rates also mean revert. \cite{boogert2008gas} used the Least Square Monte Carlo method  (initially proposed by \cite{longstaff2001valuing}  for American options)  for gas storage valuation, while assuming the following one factor model for the spot price, which is calibrated to the initial futures curve:
\begin{equation} \label{F2}
\frac{dS(t)}{S(t)}=\kappa (\mu(t)-\ln S(t))dt+\sigma dB(t), 
\end{equation}
where $B$ is a standard Brownian motion, $\mu$ is a time-dependent parameter, calibrated to the initial futures curve $(F(0,T))_{T\geq 0}$, provided by the market; the mean reversion parameter $\kappa$
and the volatility $\sigma$ are positive constants. One factor models are popular because of their simplicity. As is stated in \cite{bjerksund2008gas,henaff2013gas}, it's unrealistic to assume that one factor models can capture all characteristics of highly volatile power or natural gas prices. \cite{parsons2013quantifying} extended eq. \eqref{F2} to the two-factor mean-reverting model where the long-term mean is also a mean-reverting process. \cite{boogert2011gas} extended the application of the Least Square Monte Carlo method to multi-factor spot processes. 

The unstorability of electricity in large quantities and storage complexities of natural gas create frequent demand-supply fluctuations in the corresponding commodity markets. These disruptions lead to the spiky behavior of energy spot prices.
Therefore, price spikes have to be included in our spot price models as they carry out significant arbitrage opportunities. 

However, none of the models described above exhibit spiky instantaneous jumps of energy prices. \cite{thompson2004valuation,thompson2009natural} propose to model the risk adjusted power and natural spot prices, respectively by the following one dimensional continuous time Markov process:
\begin{equation} \label{F3}
dP = \mu_1(P,t)dt+\sigma_1(P,t)dX_1 +\sum_{k=1}^{N}\gamma_k(P,t,J_k)dq_k
\end{equation}
where $\mu, \sigma$, and the $\gamma_k$'s are any arbitrary functions and the $J_k$’s are arbitrary jump sizes with distributions $Q_k(J)$. $X_1$  and  $dq_k$'s denote the standard Brownian motion and Poisson processes respectively. Eq. \eqref{F3} covers a wide range of spot price models for natural gas and is tractable for numerical calculations. However, its jump component isn't appropriate to represent instant and mean-reverting spikes. 

\cite{benth2007non} propose a more realistic alternative by defining energy spot dynamics as a sum of Ornstein-Uhlenbeck (OU) processes with different mean-reverting speeds. The only sources of randomness of the spot prices are positive pure jump processes. The additive structure of the model makes it difficult to derive futures curves. Therefore, \cite{benth2011stochastic} considers a stochastic volatility model by \cite{barndorff2001non} for equity markets in the context of energy commodities. 
\cite{benth2013pricing} generalise this non-Gaussian approach to a multivariate case to model the dynamics of cross-commodity spot prices. 

The volatile behaviour of energy prices can differ significantly between cold and warm months. \cite{benth2008stochastic} make use of time-inhomogeneous jump processes to capture these seasonality effects by introducing geometric and arithmetic spot price dynamics in the following way:
\begin{align}  \label{F4}
\ln S(t)&=\ln\Lambda(t)+\sum_{i=1}^{m} X_i(t)+\sum_{j=1}^{n} Y_j(t), \quad \quad \text{(geometric)}\\
\intertext{and} \label{F5}
S(t)&=\Lambda(t)+\sum_{i=1}^{m} X_i(t)+\sum_{j=1}^{n} Y_j(t), \quad \quad \quad \text{(arithmetic)} \\
\intertext{where}  \label{F6}
dX_i(t)&=(\mu_i(t)-\alpha_i(t) X_i(t))dt+\sum_{k=1}^{p}\sigma_{ik}(t) dB_k(t), \\ \label{F7}
dY_j(t)&=(\delta_j(t)-\beta_j(t) Y_j(t)) dt+\eta_j(t)dL_j(t),
\end{align}
for $ i=1,\dots, m$ and $j=1,\dots,n$.  Here $B_k$ and $L_j$ are independent standard Brownian motions and independent time-inhomogeneous L\'{e}vy processes, respectively.
$\Lambda(t)$ is a deterministic, positive and continuously differentiable seasonality function, while the coefficients
  $\mu_i,\alpha_i,\delta_j,\beta_j,\eta_j$and $\sigma_{ik}$ are continuous functions of t. 

In \cite{safarov2016natural} we investigated the optimal operation problem for gas storage based on the geometric model. But for the optimisation of gas-fired power plant we will follow the arithmetic dynamics based on the justification presented in \cite{benth2008stochastic,meyer2014dynamic}.

Accurate valuation of power plants is impossible without the analysis of the multivariate dependence between electricity and gas prices. For Gaussian OU components it suffices to specify the appropriate covariance matrix. This approach is not useful when we consider non-Gaussian spot price models. Therefore 
\cite{meyer2014dynamic}  proposed to make use of L\'{e}vy copulas to model the dependence between the non-Gaussian OU components of electricity and gas spot prices. Although \cite{thompson2004valuation} develop a comprehensive stochastic control model which incorporates all the necessary operational constraints, the dependence of non-Gaussian jump components is assumed to be independent. In the actual calculations the gas prices are even assumed to be constant which is unlikely to be true under unregulated markets. 

An alternative approach is the regime-switching framework initially proposed by \cite{hamilton1990analysis}.  In a regime-switching model  the price process can randomly shift between  several regimes due to long-term demand-supply fluctuations, political instability, weather changes and other reasons. The spot price follows a distinct stochastic process within each regime. \cite{carmona2010valuation} and \cite{chen2010implications} apply a regime-switching approach to investigate stochastic control problems related to natural gas storage and implement the numerical calculation via Monte Carlo and finite difference methods, respectively. However, there is a lack of research related to the valuation of a power plant or spark-spread option under regime-switching models.

In this paper we investigate the valuation and optimal operation of gas-fired power generating facilities and refer to \cite{thompson2004valuation}for the physical operational constraints. But we assume that the underlying electricity and gas spot prices follow the L\'{e}vy regime-switching model. More precisely, at each regime the spot dynamics is defined by a special case of an arithmetic model \eqref{F7}. For the dependence of gas and electricity spikes we make use of skewed
 L\'{e}vy copulas for the reasons stated in \cite{meyer2014dynamic}. Therefore, our approach not only incorporates the most important physical characteristics of a power plant but also takes into account short-term demand-supply variations and long-term economic disruptions. 

We assume that the reader is familiar with fundamentals of the theory of L\'{e}vy
processes, as presented for example in \cite{tankov2003financial}, \cite{wim2003levy} and \cite{kyprianou2006introductory}.

To solve this stochastic control problem we employ the dynamic programming theory which gives us coupled  Hamilton-Jacobi-Bellman (HJB) equations. We then solve this system of nonlinear partial integro-differential equations (PIDE) numerically.

%
%

\section{Natural gas and electricity spot prices without regime-switching} \label{Spot prices}

Let $(\Omega,\mathcal{F},\{\mathcal{F}_t\}_{t\geq 0},\mathbb{P})$ be our complete filtered probability space. We assume that the underlying spot prices follow the arithmetic model defined by eqs. \eqref{F5} to \eqref{F7}. For simplicity, we take $m=n=p=1$. We also define the initial condition
\[
X(0)+Y(0)=S(0)-\Lambda(0).
\]
 Then we have the following dynamics for $S(t)$:

\begin{align} \label{F8}
S(t)&=\Lambda (t)+X(t)+Y(t), \\ 
\intertext{where} \label{F9}
dX(t)&=(\mu(t)-\alpha(t) X(t))dt+\sigma(t) dB(t), \\ 
 \label{F10}
dY(t)&=(\delta(t)-\beta(t) Y(t)) dt+\eta(t)dL(t),
\end{align}
We further assume that  the coefficients of eqs. \eqref{F9} and\eqref{F10} are constants such that $\mu(t)=\delta(t)=0,\alpha(t)=\beta(t)=\alpha^e, \sigma(t)=\sigma$ and $\eta(t)=1$. Then the dynamics of electricity spot prices is given by 
\begin{align} \label{F11}
S^e(t)&=\Lambda_e (t)+X^e(t)+Y^e(t), \\ 
\intertext{where} \label{F12}
dX^e(t)&=-\alpha_e X^e(t)dt+\sigma_e dB^e(t), \\ 
 \label{F13}
dY^e(t)&=-\alpha_e Y^e(t) dt+dL^e(t),
\end{align}

The underlying natural gas spot price follows the similar model:
\begin{align} \label{F14}
S^g(t)&=\Lambda_g (t)+X^g(t)+Y^g(t), \\ 
\intertext{where} \label{F15}
dX^g(t)&=-\alpha_ gX^g(t)dt+\sigma_g dB^g(t), \\ 
 \label{F16}
dY^g(t)&=-\alpha_g Y^g(t) dt+dL^g(t).
\end{align}
Note that we could take unequal mean-reversion rates for normal variations and jumps in our spot price models following  \cite{safarov2016natural}. But that would increase the dimension of our HJB equation which is derived in section \ref{Control}. We keep them equal to reduce the computational time.

In general, $L^e$ and $L^g$ can be subordinator  L\'evy processes 
\begin{equation} \notag
L^e(t)=\int_0^t\int_{ \mathbb{R_+}}zJ^e(dz,ds) \quad \quad \text{and} \quad \quad L^g(t)=\int_0^t\int_{ \mathbb{R+}}zJ^g(dz,ds)
\end{equation}
where $J^e(dz,ds)$ and $J^g(dz,ds)$ are Poisson random measures with the corresponding L\'{e}vy intensity measures $\nu^e(dz,ds)$ and $\nu^g(dz,ds)$ (see \cite{jacod2013limit}).
These jump measures have positive supports because we only consider upward price spikes. Following  \cite{meyer2014dynamic} we further assume that 
 $L^e$ and $L^g$ are compound Poisson processes with intensities $\lambda_e$ and $\lambda_g$ and jump size distributions $D_e$ and $D_g$, respectively.
 In other words, 
 \begin{align*}
 \nu^e(dz,ds)&=\lambda_e D_e(dz)ds=\nu^e(dz)ds, \\
  \nu^g(dz,ds)&=\lambda_g D_g(dz)ds=\nu^g(dz)ds.
 \end{align*}
The complete characterisation of the two-dimensional model $(S^e(t),S^g(t))$ requires the specification of the dependence risk between the spot prices $S^e$ and $S^g$. We can separate this problem in defining the multivariate distributions of 
$(B^e(t),B^g(t))$ and $(L^e(t),L^g(t))$. The former issue can be easily resolved by determining the correlation parameter $\rho$ between $B^e(1)$ and $B^g(1)$
 \[
dB^e(t)dB^g(t)=\rho dt.
\]
To specify the dependence between power and gas spot prices we will refer to \cite{meyer2014dynamic} for using positive
 L\'{e}vy copulas. This concept will be analysed in section \ref{copula}.

Before moving to the optimisation problem of a power plant we can write the electricity and gas spot price models in a more convenient SDE form as follows:
\begin{align} \notag
dS^e(t)&=d\Lambda_e (t)+dX^e(t)+dY^e(t), \\  \notag
            &=\Lambda'_e (t)dt-\alpha_e (X^e(t)+ Y^e(t))dt+\sigma_e dB^e(t)+dL^e(t) \\ \notag
            &=\Lambda'_e (t)dt-\alpha_e (S^e(t)-\Lambda_e (t))dt+\sigma_e dB^e(t)+dL^e(t) \\  \label{F17}
            &=(\bar{\Lambda}_e (t)-\alpha_e S^e(t))dt+\sigma_e dB^e(t)+dL^e(t) 
\end{align}
where $\bar{\Lambda}_e (t)=\Lambda'_e (t)+\alpha^e\Lambda_e (t)$
Similarly, we can derive the SDE for the gas spot price
\begin{equation} \label{F18}
dS^g(t)=(\bar{\Lambda}_g (t)-\alpha_g S^g(t))dt+\sigma_g dB^g(t)+dL^g(t) 
\end{equation}

Note that we can get the compensated compound Poisson processes $\tilde{L}^e(t)$ and $\tilde{L}^g(t)$ as follows:

\begin{align*}
\tilde{L}^e(t)&=\int_0^t\int_{ \mathbb{R}+}z(J^e(dz,ds)-\nu^e(dz,ds))=\int_0^t\int_{ \mathbb{R}_+}z\tilde{J}^e(dz,ds) \\ 
\tilde{L}^g(t)&=\int_0^t\int_{ \mathbb{R}+}z(J^g(dz,ds)-\nu^g(dz,ds))=\int_0^t\int_{ \mathbb{R}_+}z\tilde{J}^g(dz,ds)
\end{align*}
where $\tilde{J}^e$ and $\tilde{J}^g$ are compensated Poisson random measures.
Then eqs. \eqref{F17} and \eqref{F18} transform to
\begin{align} \label{F19}
dS^e(t)&=(\tilde{\Lambda}_e (t)-\alpha_e S^e(t))dt+\sigma_e dB^e(t)+\int_{ \mathbb{R}_+}z\tilde{J}^e(dz,dt) \\ \label{F20}
dS^g(t)&=(\tilde{\Lambda}_g (t)-\alpha_g S^g(t))dt+\sigma_g dB^g(t)+\int_{ \mathbb{R}_+}z\tilde{J}^e(dz,dt).
\end{align}
where $\tilde{\Lambda}_e (t)=\bar{\Lambda}_e (t)+\int_{ \mathbb{R}_+}z\nu^e(dz)$ and 
$\tilde{\Lambda}_g (t)=\bar{\Lambda}_g (t)+\int_{ \mathbb{R}_+}z\nu^g(dz)$.

\section{Optimal control of power generation} \label{Control}
The theoretical framework and physical constraints for
the optimal operation of a gas-fired power generator will be based on  \cite{thompson2004valuation}. Therefore, the following physical characteristics need to be incorporated in our optimisation model:
\begin{itemize}
\item The minimum generation temperature: The plant cannot
operate below a certain temperature.
\item Variable start-up times: The time required to heat the
boiler to its minimum generation level depends on the heat rate of natural gas and the current temperature.
\item  The variable output rates: The efficiency
of the plant varies non-linearly with
the amount of generated electricity.
\item The variable start-up and production costs. Increase of electricity
output is achieved by raising the boiler temperature. The fuel used for this heat increase does not generate any
additional power.
\item  Control response time lags:
Changing boiler temperature takes some time. Therefore, any decision to alter the power output will take effect after a reasonable amount
of time.
\item  The ramp rate limit: This is the minimum amount of time required for switching a unit on/off
  to avoid
thermal stress and fatigue. 
\end{itemize}
Following  \cite{thompson2004valuation} we model the boiler temperature $L$ as a state variable and incorporate all these operational constraints in its mean-reverting equation
\begin{equation} \label{F21}
dL=\eta(L,c)(\bar{L}(c)-L)dt
\end{equation}
where
$c$ is the variable controlling the amount of gas consumed per instant in time;  $\bar{L}(c)$ is the equilibrium temperature of the generator depending on $c$;
 $\eta$ is a mean-reverting speed function specific to each power plant.

 The non-linear dependence of the amount of produced electricity on the boiler temperature is given by the output function $H(L)$. We can now move to our stochastic control problem for the power plant. 

The objective of the optimisation is to find a control variable $c(S_e,S_g,L,t)$ that maximises the expected future cash flow up to maturity time $T$ and discounted at a rate $r$
\begin{equation} \label{F22}
V(S_e,S_g,L,t)=\max_{c(S_e,S_g,L,t)}\mathbb{E}\left[\int_t^T e^{-r(\tau-t)}(H(L)S_e-S_gc)d\tau\right],
\end{equation}
subject to
\begin{equation*} 
c_{min}(L)\leq c \leq c_{max}(L).
\end{equation*}
The minimum and maximum bounds on $c$ are necessary to avoid thermal stresses.

Rewriting \eqref{F22} in a similar way to \cite{thompson2004valuation} will lead to the Bellman equation:
\begin{align*}
V&=\max_c\mathbb{E}\left[\int_t^{t+dt} e^{-r(\tau-t)}(H(L)S_e-S_gc)d\tau+\int_{t+dt}^{T} e^{-r(\tau-t)}(H(L)S_e-S_gc)d\tau\right] \\  
&=\max_c\mathbb{E}\left[\int_t^{t+dt} e^{-r(\tau-t)}(H(L)S_e-S_gc)d\tau+e^{-r dt}\int_{t+dt}^{T} e^{-\rho(\tau-(t+dt))}(H(L)S_e-S_gc)d\tau\right] \\  
&=\max_c\mathbb{E}\left[\int_t^{t+dt} e^{-r(\tau-t)}(H(L)S_e-S_gc)d\tau+e^{-r dt}V(S_e+dS_e,S_g+dS_g,L+dL,t+dt)\right].
\end{align*}
 We can now apply the multidimensional It\^{o}'s formula (see \cite{tankov2003financial})
to $V(S_e,S_g,L,t)$ to expand it above in Taylor's series:
\begin{multline}
V=\max_c\mathbb{E}[(H(L)S_e-S_gc)dt+(1-r dt)[V+V_t dt+V_LdL\\ \label{F23}
+V_{S_e}dS_e+ V_{S_g} dS_g+\frac{1}{2}V_{S_e S_e}d[S_e,S_e]_t^c+\frac{1}{2}V_{S_g S_g}d[S_g,S_g]_t^c+V_{S_eS_g}d[S_e,S_g]_t^c\\ 
+\iint_{\mathbb{R}_+^2}(V(S_e+z_e,S_g+z_g,L,t)-V(S_e,S_g,L,t)-z_e V_{S_e}-z_g V_{S_g})J(dz,dt)],
\end{multline}
where $J(dz,dt)= J(dz_e,dz_g,dt)$ is a two-dimensional Poisson random measure with L\'{e}vy intensity measure $\nu(dz,dt)$ and
\begin{equation}\notag
d[S_e,S_e]_t^c=\sigma_e^2dt, \quad 
d[S_g,S_g]_t^c=\sigma_g^2dt \quad \text{and} \quad d[S_e,S_g]_t^c=\rho\sigma_e\sigma_gdt.  \label{F24}
\end{equation}

We then denote
\begin{align*}
\mu_e(S^e,t)&=\tilde{\Lambda}_e (t)-\alpha_e S^e(t)
\intertext{and}
\mu_g(S^g,t)&=\tilde{\Lambda}_g (t)-\alpha_g S^g(t)
\end{align*}
for clarity purposes.
After substituting  eqs. \eqref{F19} to \eqref{F21} and \eqref{F24} in eq. \eqref{F23} we obtain
\begin{multline} \label{F25}
V=\max_c\mathbb{E}[(H(L)S_e-S_gc)dt+(1-r dt)(V+V_t+ \\
\eta(L,c)(\bar{L}(c)-L)V_L+
\mu_e(S^e,t)V_{S_e}+\mu_g(S^g,t) V_{S_g} +\\
\frac{1}{2}\sigma_e^2V_{S_e S_e}+
\frac{1}{2}\sigma_g^2V_{S_g S_g}+\rho\sigma_e\sigma_gV_{S_eS_g})dt+\sigma_edB^e(t)+\sigma_g dB^g(t)\\ 
\iint_{\mathbb{R}_+^2}(V(S_e+z_e,S_g+z_g,L,t)-V(S_e,S_g,L,t)-z_e V_{S_e}-z_g V_{S_g})\nu(dz,dt)+\\
\iint_{\mathbb{R}_+^2}(V(S_e+z_e,S_g+z_g,L,t)-V(S_e,S_g,L,t))\tilde{J}(dz,dt))],
\end{multline}
Elimination of all terms that go to zero faster than dt and some further simplifications similar to \cite{safarov2016natural}  results in the following HJB equation
\begin{multline*} 
\max_c[V_t+\frac{1}{2}\sigma_e^2V_{S_e S_e}+
\frac{1}{2}\sigma_g^2V_{S_g S_g}+\rho\sigma_e\sigma_gV_{S_eS_g}+ 
\mu_e(S^e,t)V_{S_e}+\mu_g(S^g,t) V_{S_g} -rV+\\
\iint_{\mathbb{R}_+^2}(V(S_e+z_e,S_g+z_g,L,t)-V(S_e,S_g,L,t)-z_e V_{S_e}-z_g V_{S_g})\nu(dz,dt)+\\
H(L)S_e-S_gc+\eta(L,c)(\bar{L}(c)-L)V_L]=0
\end{multline*}
Note that the control variable $c$ only appears in the last two terms.  We also need to introduce time to maturity $\tau=T-t$. Then HJB equation above becomes
\begin{multline} \label{F26}
V_{\tau}=\frac{1}{2}\sigma_e^2V_{S_e S_e}+
\frac{1}{2}\sigma_g^2V_{S_g S_g}+\rho\sigma_e\sigma_gV_{S_eS_g}+ 
\mu_e(S^e,T-\tau)V_{S_e}+\mu_g(S^g,T-\tau) V_{S_g} -rV+\\
\iint_{\mathbb{R}_+^2}(V(S_e+z_e,S_g+z_g,L,\tau)-V(S_e,S_g,L,\tau)-z_e V_{S_e}-z_g V_{S_g})\nu(dz,d\tau)+\\
\max_c[H(L)S_e-S_gc+\eta(L,c)(\bar{L}(c)-L)V_L]=0
\end{multline}
where
\begin{equation*} 
c_{min}(L)\leq c \leq c_{max}(L).
\end{equation*}
In other words, the maximisation problem reduces to
\begin{equation}  \label{F28}
c^*(S_e,S_g,L,\tau)=\argmax_c[H(L)S_e-S_gc+\eta(L,c)(\bar{L}(c)-L)V_L]
\end{equation}
After the calculation of the optimal strategy $c^*$ we substitute it in eq. \eqref{F26} to get
\begin{multline} \label{F28}
V_{\tau}=\frac{1}{2}\sigma_e^2V_{S_e S_e}+
\frac{1}{2}\sigma_g^2V_{S_g S_g}+\rho\sigma_e\sigma_gV_{S_eS_g}+ 
\mu_e(S^e,T-\tau)V_{S_e}+\mu_g(S^g,T-\tau) V_{S_g} -rV+\\
\iint_{\mathbb{R}_+^2}(V(S_e+z_e,S_g+z_g,L,\tau)-V(S_e,S_g,L,\tau)-z_e V_{S_e}-z_g V_{S_g})\nu(dz,d\tau)+\\
H(L)S_e-S_gc^*+\eta(L,c^*)(\bar{L}(c^*)-L)V_L]
\end{multline}
Thus, once the optimal control $c^*$ is known, the nonlinear HJB equation simplifies to linear PIDE. Before starting our numerical approach to solve eq. \eqref{F28}, we analyse the two-dimensional L\'{e}vy measure $\nu$ by means of L\'{e}vy copulas in the next section.

\section{L\'{e}vy copula} \label{copula}
Before starting the analysis of the dependence of $(L^e(t),L^g(t))$ we need to go through some brief introduction on  L\'{e}vy copulas. We will refer to
 \cite{tankov2003financial,kallsen2006characterization} for the detailed background on this concept. We will mainly use the following key results in our investigation:
\begin{theorem}{(Sklar's Theorem for L\'{e}vy copulas)} \label{Tsklar}
Let $(X(t),Y(t))$ be a two-dimensional L\'{e}vy process with positive jumps that have tail integral $U(x,y)$ and marginal tail integrals $U_1(x)$ and $U_2(y)$. Then there exists a unique positive L\'{e}vy copula $F$ such that: 
\begin{equation} \label{F29}
U(x,y)=F(U_1(x),U_2(y)), \quad \forall x,y \in [0,\infty).
\end{equation}
Conversely, if $F$ is a positive L\'{e}vy copula, $X(t)$ and $Y(t)$ are one-dimensional L\'{e}vy processes with tail integrals $U_1$,$U_2$ then there exists a two-dimensional L\'{e}vy process such that its tail integral is given by eq. \eqref{F29}.
\end{theorem}
An Archimedean L\'{e}vy copula is a popular parametric L\'{e}vy copula defined as
\begin{equation*}
 F(x,y) = \phi^{-1}(\phi(x) + \phi(y)),
 \end{equation*}
 where $\phi$  is a strictly decreasing convex function with positive support such that $\phi(0) = \infty$ and $\phi (\infty)=0$. In these models $F$ is always assumed to have the symmetry property $F(x,y) = F(y,x)$. However \cite{meyer2014dynamic} argue that this symmetry property is not supported by the data. Therefore they introduce skewed Archimedean L\'{e}vy copulas as follows
\begin{equation*}
 F(x,y) = \phi^{-1}(\psi_1(y) \phi(x) +\psi_2(x) \phi(y)),
 \end{equation*}
where $\psi_1$ and $\psi_2$ are decreasing functions satisfying $\psi_1(\infty)=\psi_2(\infty)=1$, while $\phi$ is the same as above. In our calculations we will use the skewed Clayton- L\'{e}vy copula where
\[
\phi(x)=x^{-\theta}, \quad \psi_1(x)=(\alpha x^{-\beta}+1), \quad and \quad \psi_2 \equiv 1,
\]
 for $\alpha>0, \theta>0$ and $0<\beta \leq \theta+1$. More precisely,
\begin{equation} \label{F30}
F(x,y)=((\alpha y^{-\beta}+1) x^{-\theta}+y^{-\theta})^{-1/\theta}
\end{equation}

We can construct multivariate jump densities from univariate ones by using L\'{e}vy copulas. The following is the two-dimensional version of the Lemma proposed by \cite{reich2010kolmogorov}.
\begin{lemma} \label{LReich}
 Let $f\in C^2(\mathbb{R}^2)$ be a bounded function vanishing on a neighbourhood of the origin. Moreover, let  $(X(t),Y(t))$ be a two-dimensional L\'{e}vy process with L\'{e}vy measure $\nu$, L\'{e}vy copula $F$, and marginal L\'{e}vy measures $\nu_1$ and $\nu_2$. Then 
 \begin{align} \notag
\int_{\mathbb{R}^2}f(x,y)\nu(dx,dy)&=\int_{\mathbb{R}}\frac{\partial f}{\partial x}(x,0)\nu_1(dx) 
+\int_{\mathbb{R}}\frac{\partial f}{\partial y}(0,y)\nu_2(dy) \\ \label{F31}
&+\int_{\mathbb{R}^2}\frac{\partial^2 f}{\partial x\partial y}(x,y)F(U_1(x),U_2(y))dxdy.
\end{align}
\end{lemma}
We can now make use of this L\'{e}vy copula techniques to break the double integral term in eq. \eqref{F28} into parts that can be calculated via standard numerical integration techniques. First of all, let's denote the integrand as follows to reduce the size of calculations 
\begin{equation} \label{F32}
G(z_e,z_g)=V(S_e+z_e,S_g+z_g,L,\tau)-V(S_e,S_g,L,\tau)-z_e V_{S_e}-z_g V_{S_g}
\end{equation}
for fixed $S=(S_e,S_g)$ and $L$.  The application of Lemma \ref{LReich} to the integral term in eq. \eqref{F28} results in
\begin{align} \notag
\int_{\mathbb{R}_+^2}G(z_e,z_g)\nu((dz_e\times dz_g)&=\int_{\mathbb{R}_+}\frac{\partial G}{\partial z_e}(z_e,0)\nu_1((dz_e) 
+\int_{\mathbb{R}_+}\frac{\partial G}{\partial z_g}(0,z_g)\nu_2(dz_g) \\ \label{F33}
&+\int_{\mathbb{R}_+^2}\frac{\partial^2 G}{\partial z_e\partial z_g}(z_e,z_g)F(U_1(z_e),U_2(z_g))dz_e dz_g.
\end{align}

By taking the partial derivatives of $G$ in eq. \eqref{F32} we obtain
\begin{align} \notag
\frac{\partial G}{\partial z_e}(z_e,0)&=V_{S_e}(S_e+z_e,S_g,L,\tau)-V_{Se}(S_e,S_g,L,\tau) \\ \label{F34}
\frac{\partial G}{\partial z_g}(0,z_g)&=V_{S_g}(S_e,S_g+z_g,L,\tau)-V_{Sg}(S_e,S_g,L,\tau) \\ \notag
\frac{\partial^2 G}{\partial z_e\partial z_g}(z_e,z_g)&=V_{S_e S_g}(S_e+z_e,S_g+z_g,L,\tau)
\end{align}

So, after the substitution of eqs. \eqref{F33} and \eqref{F34} in eq. \eqref{F28} our HJB PIDE becomes
\begin{multline}
V_{\tau}=\frac{1}{2}\sigma_e^2V_{S_e S_e}+
\frac{1}{2}\sigma_g^2V_{S_g S_g}+\rho\sigma_e\sigma_gV_{S_eS_g}+ 
\mu_e(S^e,T-\tau)V_{S_e}+\mu_g(S^g,T-\tau) V_{S_g} -rV+\\
\int_{\mathbb{R}_+}(V_{S_e}(S_e+z_e,S_g,L,\tau)-V_{S_e}(S_e,S_g,L,\tau))\nu_e((dz_e) \\
+\int_{\mathbb{R}_+}(V_{S_g}(S_e,S_g+z_g,L,\tau)-V_{S_g}(S_e,S_g,L,\tau) )\nu_g(dz_g) \\
+\int_{\mathbb{R}_+^2}V_{S_e S_g}(S_e+z_e,S_g+z_g,L,\tau)F(U_1(z_e),U_2(z_g))dz_e dz_g \\ \label{F35}
\max_c[H(L)S_e-S_gc+\eta(L,c)(\bar{L}(c)-L)V_L]
\end{multline}
We assumed in Section \ref{Spot prices} that $L_e(t)$ and $L_g(t)$ were compound Poisson processes with intensities $\lambda_e$ and $\lambda_g$ and jump size distributions $D_e$ and $D_g$. Then the tail integrals of  $L_e(t)$ and $L_g(t)$ are

\begin{align*}
U_1(x)=&\lambda_e(1-D_e(x)), \\
U_2(x)=&\lambda_g(1-D_g(x)),
\end{align*}
respectively. If we substitute $U_1$ and $U_2$ in the formula (\ref{F30}) for the skewed Clayton- L\'{e}vy copula we get
\begin{multline}
F(U_1(z_e),U_2(z_g))=((\alpha U_2^{-\beta}+1) U_1^{-\theta}+U_2^{-\theta})^{-1/\theta}=\\ \label{F36}
\left( \left(\frac{\alpha}{\lambda_g^{\beta}(1-D_g(z_g))^{\beta}}+1\right) \frac{1}{\lambda_e^{\theta}(1-D_e(z_e))^{\theta}}+
 \frac{1}{\lambda_g^{\theta}(1-D_g(z_g))^{\theta}}\right)^{-\frac{1}{\theta}},
\end{multline}
which is a two-dimensional function depending on $z_e$ and $z_g$. We will make use of this formula later in Section \ref{Discret} in our numerical approximations.

\section{Boundary conditions} \label{Boundary}
The terminal condition for our power plant valuation is assumed to be
\begin{equation} \label{F37}
V(S_e,S_g,L,\tau=0)=0.
\end{equation}
where $\tau=T-t$.
Based on \cite{thompson2004valuation,safarov2016natural} we choose the following limits as the boundary conditions:
\begin{align} \notag
S_e \to 0 \quad \text{or} \quad S_e\to\infty \Longrightarrow \quad V_{S_e S_e} \to 0\quad   \text{and} \quad V_{S_eS_g}\to 0 \\ \label{F38}
S_g \to 0 \quad \text{or} \quad S_g \to \infty \Longrightarrow \quad V_{S_g S_g} \to 0\quad   \text{and} \quad V_{S_eS_g}\to 0
\end{align}

The domain of the HJB equation (\ref{F35}) is 
$S_e\times S_g\times L=[0,\infty)\times [0,\infty)\times [L_{\min},L_{\max}]$
But we need bounded domain for numerical calculations. So we need to restrict this unbounded domain to 
$[0, S_e^{\max},)\times [0,S_g^{\max})\times [L_{\min},L_{\max}]$.
%
Hence, for the case $S_e\to 0$ we apply $V_{S_e S_e} \to 0$ and $V_{S_eS_g}\to 0$ to simplify eq. \eqref{F35} to 
\begin{multline*}
V_{\tau}=
\frac{1}{2}\sigma_g^2V_{S_g S_g}+
\mu_e(0,T-\tau)V_{S_e}+\mu_g(S^g,T-\tau) V_{S_g} -rV+\\
\int_{\mathbb{R}}(V_{S_g}(0,S_g+z_g,L,\tau)-V_{S_g}(0,S_g,L,\tau) )\nu_g(dz_g)+ \\
\max_c[-S_gc+\eta(L,c)(\bar{L}(c)-L)V_L]
\end{multline*}
For the case $S_e \to S_e^{\max}$ we employ the first line of (\ref{F38}) again to get
\begin{multline*}
V_{\tau}=
\frac{1}{2}\sigma_g^2V_{S_g S_g}+
\mu_e(S_e^{\max},T-\tau)V_{S_e}+\mu_g(S^g,T-\tau) V_{S_g} -rV+\\
\int_{\mathbb{R}}(V_{S_g}(S_e^{\max},S_g+z_g,L,\tau)-V_{S_g}(S_e^{\max},S_g,L,\tau) )\nu_g(dz_g)+ \\
\max_c[H(L)S_e^{\max}-S_gc+\eta(L,c)(\bar{L}(c)-L)V_L]
\end{multline*}
We can obtain the boundary conditions for the limits $S_g\to 0$ and $S_g \to S_g^{\max}$ in a similar way.
 To conclude our analysis of boundary conditions we need to construct the equations corresponding to 4 corner points: $(S_e,S_g) \to (0,0)$, 
 $(S_e,S_g) \to (0,S_g^{\max})$, $(S_e,S_g) \to (S_e^{\max},0)$ and 
 $(S_e,S_g) \to (S_e^{\max},S_g^{\max})$.
For the first two corners we have
\begin{multline*}
V_{\tau}=
\mu_e(0,T-\tau)V_{S_e}+\mu_g(0,T-\tau) V_{S_g} -rV+
\max_c[\eta(L,c)(\bar{L}(c)-L)V_L]
\end{multline*}
and
\begin{multline*}
V_{\tau}=
\mu_e(0,T-\tau)V_{S_e}+\mu_g(S_g^{\max},T-\tau) V_{S_g} -rV+\\
\max_c[-S_g^{\max}c+\eta(L,c)(\bar{L}(c)-L)V_L]
\end{multline*}
Similarly, we can derive the equations for the other two corners.

\section{Discretisation scheme} \label{Discret}

$L^e$ and $L^g$ are finite activity L\'{e}vy processes where
$\nu_e(\mathbb{R}_{+})=\lambda_e$ and $\nu_g(\mathbb{R}_{+})=\lambda_g$.
%
Therefore, we can rewrite the single integral terms of eq. \eqref{F35} as
\begin{align} \notag
\int_{\mathbb{R}_{+}}(V_{S_e}(S_e+z_e,S_g,L,\tau)-V_{S_e}(S_e,S_g,L,\tau))\nu_e(dz_e) =\\ \label{F39}
\int_{\mathbb{R}_{+}}V_{S_e}(S_e+z_e,S_g,L,\tau)\nu_e(dz_e) -\lambda_{e}V_{S_e}(S_e,S_g,L,\tau)
\end{align}
and
\begin{align} \notag
\int_{\mathbb{R}_+}(V_{S_g}(S_e,S_g+z_g,L,\tau)-V_{S_g}(S_e,S_g,L,\tau) )\nu_g(dz_g)=\\ \label{F40}
\int_{\mathbb{R}_+}V_{S_g}(S_e,S_g+z_g,L,\tau)\nu_g(dz_g)-\lambda_g V_{S_g}(S_e,S_g,L,\tau) 
\end{align}
If we substitute eqs. \eqref{F39} and \eqref{F40} in eq. \eqref{F35} the coefficients of partial derivatives
 $V_{S_e}$ and $V_{S_g}$ modify to
\begin{align*}
\bar{\mu}_e(S^e,t)&=\bar{\Lambda}_e (t)-\alpha_e S^e(t)-\lambda_e\\
\bar{\mu}_g(S^g,t)&=\bar{\Lambda}_g (t)-\alpha_g S^g(t)-\lambda_g
\end{align*}
and the HJB equation (\ref{F35}) can be written as
\begin{multline} \label{F41}
V_{\tau}=\mathcal{D}V+\mathcal{H}_eV+\mathcal{H}_gV+\mathcal{H}_{eg}V+\\
\max_c[H(L)S_e-S_gc+\eta(L,c)(\bar{L}(c)-L)V_L]
\end{multline}
where
\begin{align*}
\mathcal{D}V=&\frac{1}{2}\sigma_e^2V_{S_e S_e}+
\frac{1}{2}\sigma_g^2V_{S_g S_g}+\rho\sigma_e\sigma_gV_{S_eS_g}+ \\
&\mu_e(S^e,T-\tau)V_{S_e}+\mu_g(S^g,T-\tau) V_{S_g} -rV \\
\mathcal{H}_eV=&\int_{\mathbb{R}_{+}}V_{S_e}(S_e+z_e,S_g,L,\tau)\nu_e((dz_e) ,\\
\mathcal{H}_gV=&\int_{\mathbb{R}_+}V_{S_g}(S_e,S_g+z_g,L,\tau)\nu_g(dz_g)
\intertext{and}
\mathcal{H}_{eg}V=&\int_{\mathbb{R}_+^2}V_{S_e S_g}(S_e+z_e,S_g+z_g,L,\tau)F(U_1(z_e),U_2(z_g))dz_e dz_g
\end{align*}
Note that we dropped the bar signs from our coefficients $\bar{\mu}_e$ and $\bar{\mu}$ for simplicity.
We will apply the explicit finite difference scheme proposed in \cite{cont2005finite,cont2005integro} to deal with the diffusion $\mathcal{D}V$ and single integral terms $\mathcal{H}_eV$ and $\mathcal{H}_gV$. The discretisation of $\mathcal{H}_{eg}V$ and optimisation terms requires further analysis. 

First of all, we need to discretise the domain of the HJB equation by the following grid:
\begin{align*}
\tau_{n}&=n\Delta\tau, \quad n=0,\dots,M, \quad  \Delta \tau=\frac{T}{M}, \\
S^e_{i}&=i\Delta S^e, \quad i=0,\dots,N_e, \quad  \Delta S^e=\frac{S^e_{\max}}{N_e}, \\
S^g_j&=j\Delta S^g, \quad j=0,\dots,N_g, \quad  \Delta S^g=\frac{S^g_{\max}}{N_g}, \\
L_u&=L_{\min}+u\Delta L, \quad u=0,\dots,L, \quad  \Delta L=\frac{L_{max}-L_{\min}}{N_L}.
\end{align*}
So, $V(S^e_i,S^g_j,L_u,\tau_n)=V(i\Delta S^e,j \Delta S^g, u\Delta L, n\Delta \tau)$ is the exact solution of the eq. \eqref{F35} at the node $(S^e_i,S^g_j,L_u,\tau_n)$, while $V^n_{i,j,u}$ stands for the approximate solution at the same node. 

For the diffusion/PDE part we follow the steps discussed in \cite{safarov2016natural} use the explicit method discussed in \cite{safarov2016natural} and apply similar corrector term to get rid off the monotonicity issue. In other words,
\begin{multline} \label{F42}
\mathcal{D}V^n_{i,j,u}=\frac{\sigma_e^2}{2 \Delta S_e^2}(V^{n}_{i+1,j,u}-2V^n_{i,j,u}+V^n_{i-1,j,u})+
\frac{\sigma_g^2}{2 \Delta S_g^2}(V^{n}_{i,j+1,u}-2V^n_{i,j,u}+V^n_{i,j-1,u})+\\
\frac{\rho\sigma_e\sigma_g}{4 \Delta S_e\Delta S_g}(V^{n}_{i+1,j+1,u}+V^{n}_{i-1,j-1,u}-V^n_{i,j+1,u}-V^n_{i,j-1,u}-V^n_{i+1,j,u}-V^n_{i-1,j,u}+2V^n_{i,j,u})+ \\
\frac{\mu_e(S^e_i,T-\tau_n)}{\Delta S_e}(V^{n}_{i+1,j,u}-V^n_{i,j,u})+\frac{\mu_g(S^g_j,T-\tau_n)}{\Delta S_g}
(V^{n}_{i,j+1,u}-V^n_{i,j,u}) -rV^n_{i,j,u}
\end{multline}
assuming that $\mu_e\geq0$ and $\mu_g\geq 0$. We will use the downwind scheme for the cases $\mu_e\leq0$ and $\mu_g\leq 0$.

For the integral terms  $\mathcal{H}_eV$ and $\mathcal{H}_gV$ we first need to restrict the unbounded integration domains by setting finite upper bounds $B_e$ and $B_e$ to truncate the larger jumps. We employ the trapezoidal quadrature method in order to get the numerical approximations for these integrals and take 
$\Delta z_e=\Delta S_e$ and $\Delta z_g=\Delta S_g$. We also choose positive integers $K_e$ and $K_g$ such that 
 $[0,B_e] \subset [0, (K_e+\frac{1}{2})\Delta z_e]$ and $[0,B_g] \subset [0, (K_g+\frac{1}{2})\Delta z_g]$, respectively. By using the central difference discretisation for the partial derivatives $V_{S_e}$ and $V_{S_g}$ we get
\begin{align} \notag
\mathcal{H}_eV^n_{i,j,u}=&\int_0^{B_e} V_{S_e}(S^e_i+z_e,S^g_j,L_u,\tau_n)\nu_e((dz_e)\\ \notag
                                         =&\frac{1}{2\Delta S_e}\int_0^{B_e} (V(S^e_{i+1}+z_e,S^g_j,L_u,\tau_n)-V(S^e_{i-1}+z_e,S^g_j,L_u,\tau_n))\nu_e((dz_e) \\ \label{F43}
\approx &
\frac{1}{2\Delta S^e}(\sum_{k=0}^{K_e}\nu^e_k V^n_{i+k+1,j,u}-\sum_{k=0}^{K_e}\nu^e_k V^n_{i+k-1,j,u})
\end{align}
and
\begin{align} \notag
\mathcal{H}_gV^n_{i,j,u}=&\int_0^{B_g} V_{S_g}(S^e_i,S^g_j+z_g,L_u,\tau_n)\nu_g((dz_g) \\ \notag
 =&\frac{1}{2\Delta S_g}\int_0^{B_g} (V(S^e_{i},S^g_{j+1}+z_g,L_u,\tau_n)-V(S^e_{i},S^g_{j-1}+z_g,L_u,\tau_n))\nu_g((dz_e) \\ \label{F43}
\approx &
\frac{1}{2\Delta S_g}(\sum_{k=0}^{K_g}\nu^g_k V^n_{i,j+k+1,u}-\sum_{k=0}^{K_g}\nu^g_k V^n_{i,j+k-1,u})
\end{align}
where
\[
\nu_k^e=\int_{(k-\frac{1}{2})\Delta S_e}^{(k+\frac{1}{2})\Delta S_e} \nu_e(dz_e) \quad \text{and} \quad 
\nu_k^g=\int_{(k-\frac{1}{2})\Delta S_g}^{(k+\frac{1}{2})\Delta S_g} \nu_g(dz_g)
\]
We can now go through the numerical calculation of the double integral 
\[
\mathcal{H}_{eg}V=\int_0^{B_e}\int_0^{B_g}V_{S_e S_g}(S_e+z_e,S_g+z_g,L,\tau)F(U_1(z_e),U_2(z_g))dz_e dz_g 
\]
First of all, denote $\bar{F}(z_e,z_g)=F(U_1(z_e),U_2(z_g))$ for clarity purposes. Then apply the central difference formula to approximate the cross-derivative  $V_{S_e S_g}$ at the node $(S^e_i+z_e,S^g_j+z_g,L_u,\tau_n)$:
\begin{multline*}
V_{S_e S_g}(S^e_i+z_e,S^g_j+z_g,L_u,\tau_n) \approx 
\frac{1}{4 \Delta S_e\Delta S_g}  (V(S^e_{i+1}+z_e,S^g_{j+1}+z_g,L_u,\tau_n)-\\
V(S^e_{i+1}+z_e,S^g_{j-1}+z_g,L_u,\tau_n)-V(S^e_{i-1}+z_e,S^g_{j+1}+z_g,L_u,\tau_n)\\
+V(S^e_{i-1}+z_e,S^g_{j-1}+z_g,L_u,\tau_n))
\end{multline*}
By taking the double integral of both sides above we get
\begin{multline}
\mathcal{H}_{eg}V \approx 
\frac{1}{4 \Delta S_e\Delta S_g} \int_0^{B_e}\int_0^{B_g} (V(S^e_{i+1}+z_e,S^g_{j+1}+z_g,L_u,\tau_n)-\\
V(S^e_{i+1}+z_e,S^g_{j-1}+z_g,L_u,\tau_n) 
-V(S^e_{i-1}+z_e,S^g_{j+1}+z_g,L_u,\tau_n)+\\ \label{F45}
V(S^e_{i-1}+
z_e,S^g_{j-1}+z_g,L_u,\tau_n))\bar{F}(z_e,z_g)dz_e dz_g
\end{multline}
Thus, we need to integrate 4 terms over $[0,B)\times[0,B)$ in eq. \eqref{F45} in order to evaluate $\mathcal{H}_{eg}V$. For this purpose we will make use of two-dimensional trapezoidal rule explained below (see \cite{davis2007methods}).

Let $f(x,y)$ be given over the rectangle 
$\{(x,y): 0\leq x \leq a, 
 0 \leq y \leq b\}$. Assume that the intervals $[0,a]$ and $[0,b]$ are equally subdivided into $K_1$ and $K_2$ subintervals with widths $h_1=\frac{a}{K_1}$ and $h_2=\frac{b}{K_2}$, respectively. The sample points $x_i$ and $y_j$ are defined as
$x_i=ih_1$ and $y_j=jh_2$ where $i=0,\dots K_1$ and $j=0,\dots K_2$. Then the two-dimensional trapezoidal rule can be formulated as follows
\[
\int_0^a\int_0^b f(x,y)dxdy\approx \sum_{i=0}^{K_1} \sum_{j=0}^{K_2} \omega_{ij} f(x_i,y_j)h_1h_2
\]
where 
\[
\omega_{ij}=
\begin{cases}
1 \quad &\text{for}  \quad  \quad   0<i<K_1 \quad \text{and} \quad 0<j<K_2 \quad   \quad \quad     \text{interior} \\
\frac{1}{2} \quad & \text{for} \quad \quad  (0,j) \quad \text{or} \quad (i,0) \quad \quad \quad \quad \quad \quad \quad \quad \quad\text{exterior}  \\
\frac{1}{4} \quad & \text{fpr}                 \quad   \quad   (i,j) \in \{(0,0),(N_B,N_B)\}  \quad   \quad \quad \quad    \quad \text{corners.}
\end{cases}
\]
We apply this integration rule to approximate the four terms on the right hand side (RHS) of eq. \eqref{F45}. Let's first consider the first term
\[
f(z_e,z_g)=V(S^e_{i+1}+z_e,S^g_{j+1}+z_g,L_u,\tau_n)\bar{F}(z_e,z_g)
\]
This is a two-dimensional function depending on $z_e$ and $z_g$. So, we can apply the 2d trapezoidal rule to get
\begin{multline*}
\int_0^{B_1}\int_0^{B_2}V(S^e_{i+1}+z_e,S^g_{j+1}+z_g,L_u,\tau_n)\bar{F}(z_e,z_g)dz_edz_g \approx \\
\sum_{k_1=0}^{K_1} \sum_{k_2=0}^{K_2} V(S^e_{i+k_1+1},S^g_{j+k_2+1},L_u,\tau_n)\bar{F}(S^e_{k_1},S^g_{k_2}) h_1h_2\omega_{k_1 k_2}
\end{multline*}
We can similarly calculate the other three terms.

Hence, after denoting the approximate value of $c(S^e_i,S^g_j,L_u,\tau_n)$ by $\Omega^n_{i,j,u}$ our explicit scheme becomes
\begin{multline}  \label{F46}
\frac{V^{n+1}_{i,j,u}-V^n_{i,j,u}}{\Delta \tau}=D V^n_{i,j,u}+H_eV^n_{i,j,u}+H_gV^n_{i,j,u}+H_{eg}V^n_{i,j,u} \\ 
\max_{\Omega^n_{i,j,u} \in C(I)}\left\{H(L_u)S^e_i-S^g_j\Omega^n_{i,j,u}+\eta(L_u,\Omega^n_{i,j,u})(\bar{L}(\Omega^n_{i,j,u})-L_u)\frac{V^{n}_{i,j,u+1}-V^n_{i,j,u}}{\Delta L}\right\}
\end{multline}
where $D,H_e,H_g$ and $ H_{eg}$ are discretisation operators corresponding to  $\mathcal{D},\mathcal{H}_e,\mathcal{H}_g$ and $\mathcal{H}_{eg}$, respectively. Note that when $\eta(L_u,\Omega^n_{i,j,u})(\bar{L}(\Omega^n_{i,j,u})-L_u)<0$ we will discretise $V_L$ by $({V^{n}_{i,j,u}-V^n_{i,j,u-1}})/{\Delta L}$. 

 We know from the initial condition that $V^{0}_{i,j,u}=0$. So, starting from time $\tau_0=0$ we first solve the maximization problem
\begin{equation*} 
\hat{\Omega}^n_{i,j}=\argmax_{\Omega^n_{i,j,u} \in C(I)}\left\{H(L_u)S^e_i-S^g_j\Omega^n_{i,j,u}+\eta(L_u,\Omega^n_{i,j,u})(\bar{L}(\Omega^n_{i,j,u})-L_u)\frac{V^{n}_{i,j,u+1}-V^n_{i,j,u}}{\Delta L}\right\},
\end{equation*}
then substitute this optimal strategy $\hat{\Omega}^n_{i,j}$ in \eqref{F46} to get
\begin{multline}  \label{F47}
V^{n+1}_{i,j,u}=V^n_{i,j,u}+\Delta \tau (D V^n_{i,j,u}+ H_e V^n_{i,j,u}+ H_g V^n_{i,j,u}+ H_{eg} V^n_{i,j,u}) \\  
\Delta \tau \left\{H(L_u)S^e_i-S^g_j\hat{\Omega}^n_{i,j,u}+\eta(L_u,\hat{\Omega}^n_{i,j,u})(\bar{L}\hat{\Omega}^n_{i,j,u})-L_u)\frac{V^{n}_{i,j,u+1}-V^n_{i,j,u}}{\Delta L}\right\}
\end{multline}
where $n=0,1,...,M-1$.

\section{Hypothetical gas-fired power plant example}
 As we mentioned earlier, we want to model the power output as a function
of heat $L$. We specifically refer to \cite{thompson2004valuation} in assuming 
\[
H(L)=
\begin{cases}
0 \quad &\text{for}    \quad \quad L<300   \\
\frac{5}{6}L-100 \quad & \text{for} \quad \quad 300 \leq L \leq 600
\end{cases}
\]
Hence, the minimum and maximum operating temperatures for electricity generation are $300 ^{\circ}$C and $600 ^{\circ}$C, respectively which correspond to 150 MW and 400 MW output levels. As we are dealing with a dynamic model, we also need the equilibrium temperature for the boiler that depends on the amount of burned gas $c$:
\begin{equation*}
\bar{L}(c)=b_0-b_1(c-b_2)^2,
\end{equation*}
where $(b_0,b_1,b_2)=(650,0.00003571,4200)$.
By assuming $\eta(L,c)=\eta=0.1$ in eq. \eqref{F21} we get the following dynamics for the temperature
\begin{equation} \label{F48}
\frac{dL}{dt}=0.1(\bar{L}(c)-L),
\end{equation}
where $0 \leq c\leq 3017$. We then impose a ramp rate restriction 
\begin{equation*} 
\left|\frac{dL}{dt}\right| \leq 15.
\end{equation*}
proposed by \cite{thompson2004valuation}.
If we combine this restriction with \eqref{F48}, we can update the lower and upper boundaries for the control variable $c$ in the following way
\begin{align*}
\bar{c}_{\min}(L) &\leq c \leq \bar{c}_{\max}(L),  \quad  \quad 20 \leq L \leq 600\\
\intertext{where}
\bar{c}_{\min}(L)&=\max\left(0,4200-100 \sqrt{\frac{800-L}{0.3571}}\right), \\
\bar{c}_{\max}(L)&=\min\left(3017,4200-100 \sqrt{\frac{500-L}{0.3571}}\right)
\end{align*}
Note the discontinuity of $H(L)$ at $L=300$ causes some oscillation problems related to the discretisation of the partial derivative $V_L$. with the smothness of solutions. We will follow \cite{thompson2004valuation} in making use of  minmod slope limiters to get rid off this smoothness issue. The detailed explanation regarding slope or flux limiters is provided by 
\cite{leveque1992numerical}.

\subsection{Slope limiter}
The following calculations do not affect the diffusion and integration parts of the eq. \eqref{F35}. Therefore, we can omit them in order to simplify our PIDE to
\begin{equation} \label{F49}
V_{\tau}=
H(L)S_e-S_gc^*+\eta(L,c^*)(\bar{L}(c^*)-L)V_L]
\end{equation}
If we denote $a(L)=\eta(L,c^*)(\bar{L}(c^*)-L)$ in eq. \eqref{F49} then we can see that 
\begin{equation} \label{F50}
V_{\tau}=
H(L)S_e-S_gc^*+a(L)V_L]
\end{equation}
is an advection equation with variable speed $a(L)$. According to eq. \eqref{F47}, we find $c^*$ before applying explicit scheme to discretise \eqref{F50}. Therefore, we can extend the MUSCL scheme (Monotonic Upwind-Centered Scheme for Conservation Laws) for advection equations with constant speed to cover eq. \eqref{F50} as well. So,
\begin{multline}
V^{n+1}_u=V^n_u+\Delta \tau (H(L_u)S^e_i-S^g_j\hat{\Omega}^n_{u})-\zeta (V^n_{u_1}-V^n_{u_1-1})-\\
\frac{1}{2}\zeta(\sign(\zeta)-\zeta)\Delta L (\sigma_{u_1}-\sigma_{i_1-1}),
\end{multline}
where $\zeta=\frac{\Delta \tau a(L_u)}{\Delta L}$ and
\[
u_1=
\begin{cases}
u \quad &\text{if} \quad \zeta>0\\
u+1 \quad &\text{if} \quad \zeta \leq 0
\end{cases}
\]
$\sigma^n_{u}$ is some slope that needs to be defined in order to reduce the numerical oscillations.
 The oscillations in a solution are  measured by the total variation (TV) of $V$:
\[
TV(V)=\sum_{u=1}^{N}|V_u-V_{u-1}|
\]
The increase of TV leads to the development of oscillations in a solution. Therefore, a numerical method is total variation diminishing (TVD) if 
\[
TV(V^{n+1})\leq TV(V^n)
\]
We need to limit the slope $\sigma^n_u$ of the MUSCL scheme to make it TVD. One of such ways is the minmod slope limiter:
\begin{align*}
\sigma^n_u&=\minmod\left(\frac{V^{n}_{u}-V^n_{u-1}}{\Delta L},\frac{V^{n}_{u+1}-V^n_{u}}{\Delta L} \right)
\intertext{where}
\minmod(a,b)&=
\begin{cases}
a \quad &\text{if}    \quad \quad  |a|<|b| \quad \text{and}  \quad ab>0 \\
b \quad & \text{if}  \quad \quad  |a|>|b| \quad \text{and}  \quad ab>0 \\
0 \quad & \text{if}   \quad   \quad  ab\leq 0
\end{cases} \\ 
&=\frac{1}{2}(\sign(a)+\sign(b))\min(|a|,|b|).
\end{align*}
Thus, after improving the numerical algorithm \eqref{F47} via the minmod slope limiter technique, we can get benefit from the accuracy of second order differencing scheme while avoiding numerical oscillations. 

\subsection{Numerical results} \label{Secr0}
\begin{figure}[h!]
	\centering
	\includegraphics[width=0.8\textwidth]{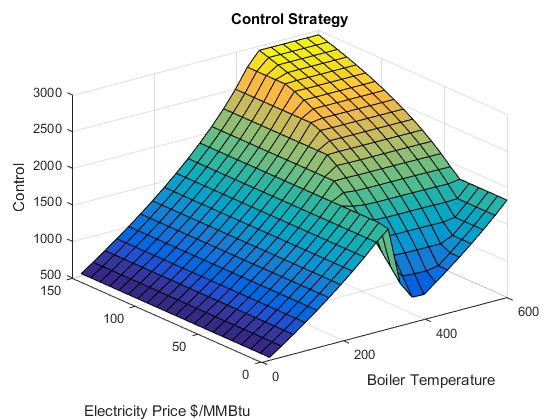}
	\caption{Control strategy surface for fixed gas price $S_g=3.5$.}
	\label{Fig2}
\end{figure}

We first consider the framework similar to the one presented in \cite{thompson2004valuation}. More precisely, we assume that the natural gas spot price is constant $S_g=3.5 MMBtu$ and the input parameters are 
\[
(\alpha_e,\sigma_e,\lambda_e,T,r)=(.1,.12,.1,200~ hours,.05).
\]
Moreover, power spot price jump sizes are normally distributed, $\mathcal{N}(700,100^2)$. The deterministic seasonality function $\Lambda_e$ is given as
\[
\Lambda_e(t)=15 \sin((2\pi t-15.4\pi)/24)+27,
\]
The optimal operating strategy surface for the gas-fired power plant is presented in Figure \ref{Fig2}. For boiler temperatures below 300$^{\circ}$C the optimal strategy is to burn maximum amount of fuel within physical limitations in order to keep the plant on line. For low electricity spot prices gas consumption decreases when the temperature is above the minimum operation level to minimise the cost. But for higher power prices the optimal control of gas consumption increases nonlinearly w.r.t. spot prices and temperature. We can see from Figure \ref{Fig2} that in the case of low electricity spot prices the power plant burns more fuel for very high temperatures. This strategy is explained by the ramp rate restriction. In other words, we can't shut down the plant immediately when the boiler is close to its maximum temperature limit. It will take some time for the unit to cool and we have to burn the required amount of gas during that time period. 

\begin{figure}[h!]
	\centering
	\includegraphics[width=0.8\textwidth]{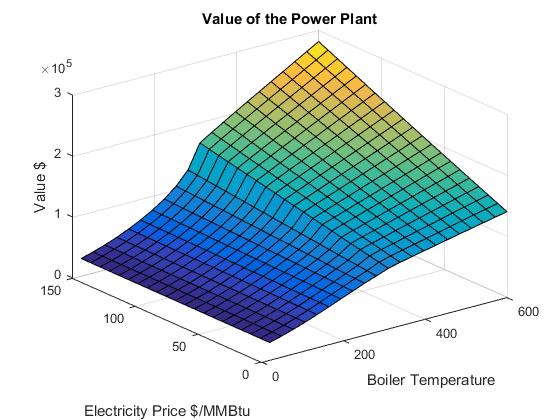}
	\caption{Value surface for fixed gas price $S_g=3.5$.}
	\label{Fig3}
\end{figure}

The Figure \ref{Fig3} shows the value or expected cash flow surface 
of power generating facility. We can see that the value surface does not depend on electricity spot prices when the boiler temperature is below the operating level. The plant is not reacting to spot price changes because there is no guarantee that by the time we increase the temperature to 300$^{\circ}$C spot prices will remain the same. Beyond 300$^{\circ}$C the plant starts to generate electricity which explains the non-smooth behaviour of the value surface at $L=300$. Beyond this line both higher power prices and boiler temperatures contribute to larger values of the power generator.
Thus, our optimal control and value surfaces look similar to the figures given by \cite{thompson2004valuation}. Unfortunately, we can't conduct more accurate comparison due to some missing parameter specifications in the original paper. 

We can now investigate the numerical calculations for the underlying spot price models \eqref{F11} and \eqref{F14}. The parameters for our single regime model are given in Table \ref{T1}.
 \begin{table}
 	\begin{center}
 		\begin{tabular}{| l | l | l | l |}
 			\hline
 			Parameter & Value & Parameter & Value \\ \hline
 			$\alpha_e$ & .1   & $\lambda_e$ & .1\\ \hline
 				$\alpha_g$ & .23  &   $\lambda_g$ & .4\\ \hline
 		 $\sigma_e$ & .11  & $T$ & 200 hours \\
 			\hline
 			 $\sigma_g$ & .09  & $\rho$ & .15  \\\hline
 		\end{tabular}
 	\end{center}
 		\caption{Input parameters for the single regime power plant model}
 	\label{T1}
 \end{table}
We assume that the risk-free interest rate is the same, $r=0.05$.  As the jump size distributions $D_e$ and $D_g$ we take inverse Gaussian distributions $IG(\mu,\lambda)$ where $\mu>0$ and $\lambda>0$ are mean and shape parameters, respectively. We can use the parameter values $(.60,.56)$ and $(.54,.32)$ for the corresponding distributions of power and gas price jumps. 

\begin{figure}
	\centering
	\begin{subfigure}[b]{0.55\textwidth}
		\includegraphics[width=\textwidth]{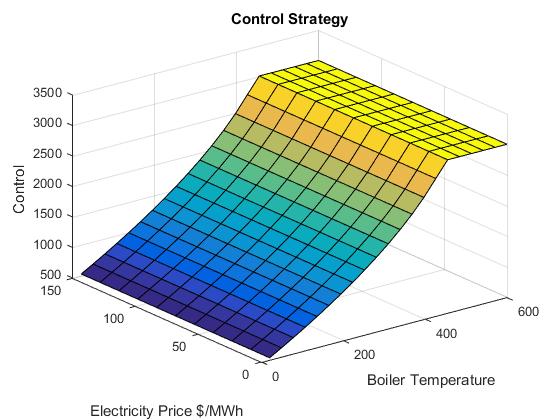}
		\caption{$S_g=0$}
		\label{F4a}
	\end{subfigure}%
	~ 
	\begin{subfigure}[b]{0.55\textwidth}
		\includegraphics[width=\textwidth]{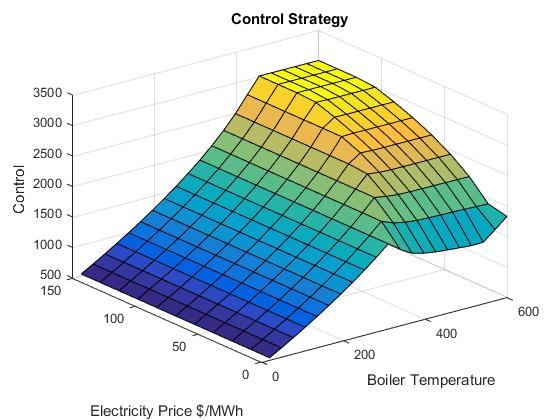}
		\caption{$S_g=10$}
		\label{F4b}
	\end{subfigure}
	
	\begin{subfigure}[b]{0.55\textwidth}
		\includegraphics[width=\textwidth]{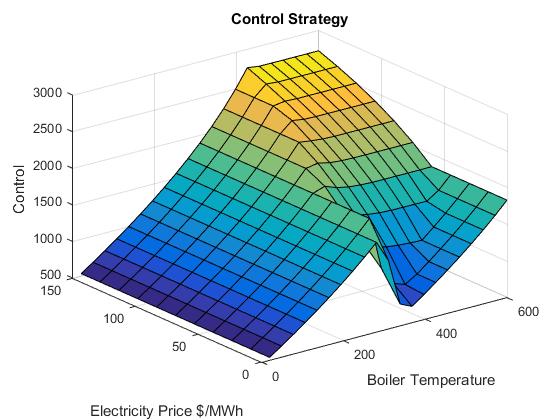}
		\caption{$S_g=14$}
		\label{F4c}
	\end{subfigure}%
	~ 
	\begin{subfigure}[b]{0.55\textwidth}
		\includegraphics[width=\textwidth]{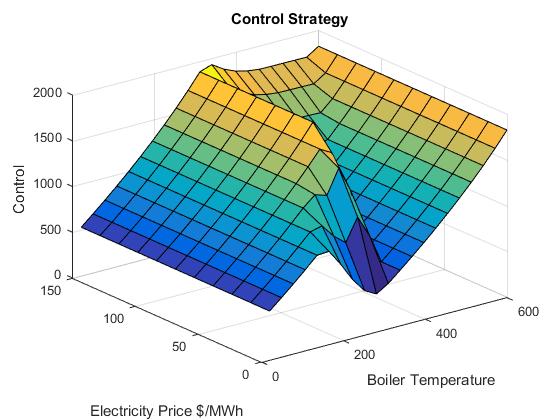}
		\caption{$S_g=20$}
		\label{F4d}
	\end{subfigure}
	\caption{Operating strategy for the single regime model w.r.t. electricity spot price and boiler temperature for different gas prices}\label{Fig4}
\end{figure}

Meanwhile, the deterministic seasonality terms $\Lambda_e$ and $\Lambda_g$ will be defined as follows
\begin{align*}
\Lambda_e(t)&=15 \sin((2\pi t-15.4\pi)/24)+27, \\
\Lambda_g(t)&=0.6 \cos(2\pi (t-18\pi)/24)+2.7
\end{align*}
Figure \ref{Fig4} compares optimal operating surfaces w.r.t. power prices and boiler temperature for various gas spot prices. When $S_g\to 0$, the generation costs are so small that the plant burns gas as much as possible at all levels of power prices and temperatures (Figure \ref{F4a}). Figure \ref{F4b} depicts the case when $S_g=10$\$/MMBtu. We can see that the surface looks more like in \ref{Fig2}. For higher levels of natural gas spot prices the region of intensive fuel consumption diminishes (Figure \ref{F4c} and \ref{F4d}).

\begin{figure}
	\centering
	\begin{subfigure}[b]{0.55\textwidth}
		\includegraphics[width=\textwidth]{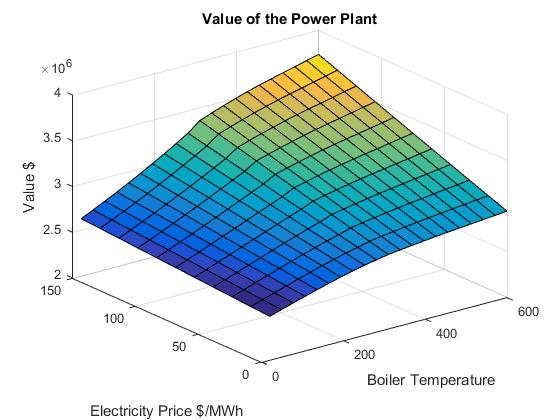}
		\caption{$S_g=0$}
		\label{F5a}
	\end{subfigure}%
	~ 
	\begin{subfigure}[b]{0.55\textwidth}
		\includegraphics[width=\textwidth]{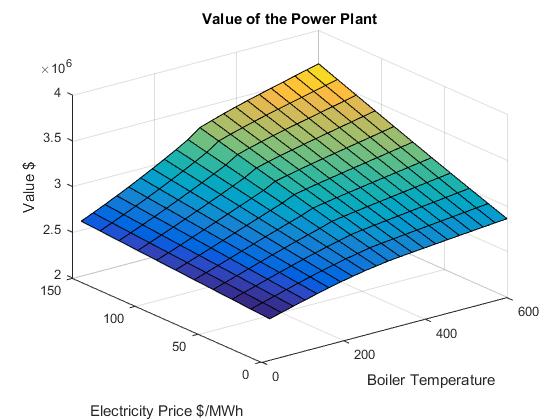}
		\caption{$S_g=10$}
		\label{F5b}
	\end{subfigure}
	
	~ 
	\begin{subfigure}[b]{0.55\textwidth}
		\includegraphics[width=\textwidth]{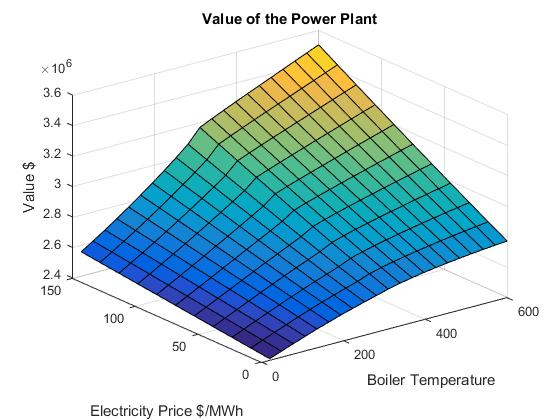}
		\caption{$S_g=20$}
		\label{F5c}
	\end{subfigure}
 \caption{Value surface for the single regime model w.r.t. electricity spot price and boiler temperature for different gas prices}\label{Fig5}
\end{figure}

Value surfaces w.r.t. power spot price and boiler temperature for different gas prices are depicted in Figure \ref{Fig5}. The shapes of these surfaces are almost the same. As gas prices increase, the surfaces shift downwards. But when $S_g$ is too high, the power plant value becomes dependant on power spot prices even for low temperature levels (Figure \ref{F5c}).

\begin{figure}
	\centering
	\begin{subfigure}{0.32\textwidth}
		\centering
		\includegraphics[width=\textwidth]{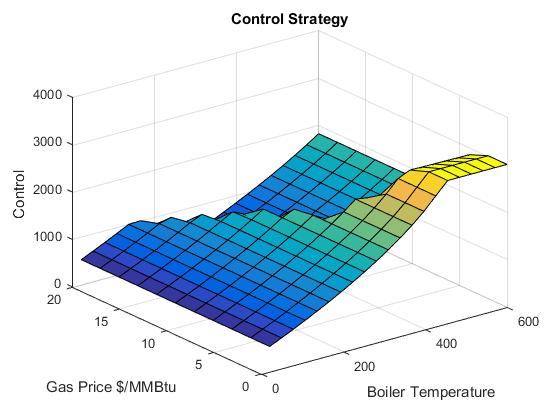}
		\caption{$S_e=0$}
		\label{F6a}
	\end{subfigure}%
	~ 
	\begin{subfigure}{0.32\textwidth}
		\centering
		\includegraphics[width=\textwidth]{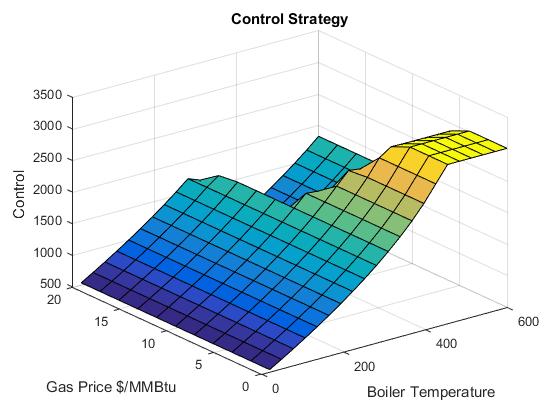}
		\caption{$S_e=60$}
		\label{F6b}
	\end{subfigure}
	\begin{subfigure}{0.32\textwidth}
		\centering
		\includegraphics[width=\textwidth]{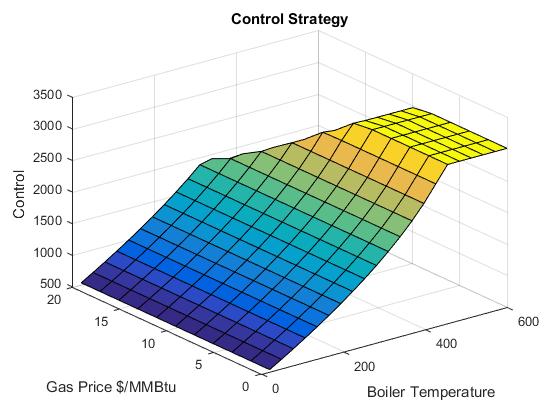}
		\caption{$S_e=150$}
		\label{F6c}
	\end{subfigure}%
	
	\begin{subfigure}{0.32\textwidth}
		\centering
		\includegraphics[width=\textwidth]{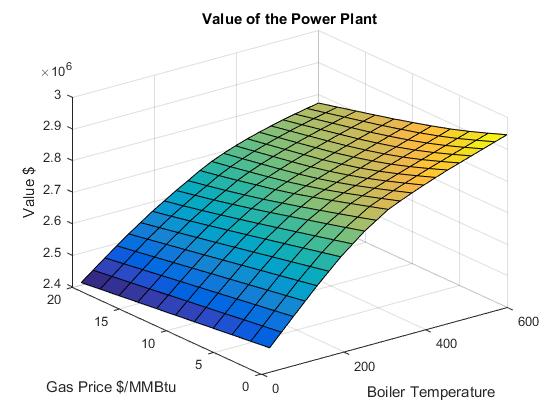}
		\caption{$S_e=0$}
		\label{F6d}
	\end{subfigure}%
	~ 
	\begin{subfigure}{0.32\textwidth}
		\centering
		\includegraphics[width=\textwidth]{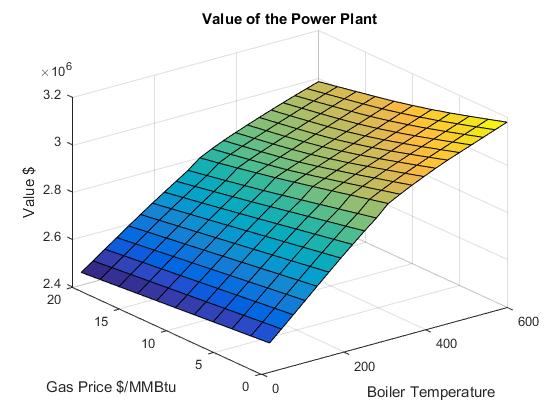}
		\caption{$S_e=60$}
		\label{F6e}
	\end{subfigure}
	\begin{subfigure}{0.32\textwidth}
		\centering
		\includegraphics[width=\textwidth]{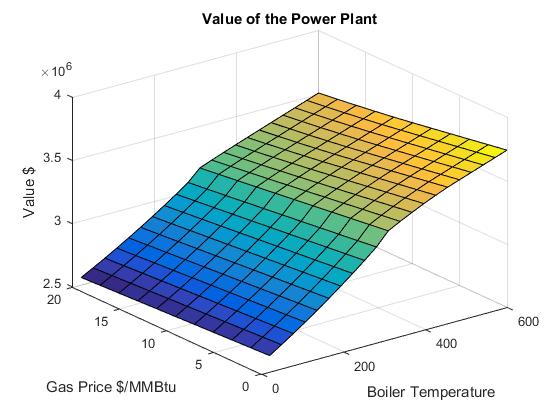}
		\caption{$S_e=150$}
		\label{F6f}
	\end{subfigure}%
	\caption{Operating strategy (a)-(c) and value surfaces (d)-(f) for a single regime model w.r.t. gas spot price and boiler temperature for different electricity prices}\label{Fig6}
\end{figure}

In Figure \ref{Fig6} we can observe optimal control and corresponding value surfaces w.r.t. gas spot price and boiler temperature for different electricity prices. Figures (a)-(c) imply that the higher is $S_e$, the larger is the maximum consumption region. For very low values of electricity the plant minimises its fuel cost even before reaching 300$^{\circ}$C. But when the power prices are too high the plant can keep burning gas at full capacity beyond 300$^{\circ}$C as well. Running the plant in the presence of extremely expensive fuel prices when the boiler temperature is too high becomes unbearable. Therefore, the optimal strategy is to reduce gas consumption to minimum. 

We can see from Figures \ref{Fig6}(d)-(f) that value surfaces shift upwards when $S_e$ increases. Meanwhile, the value surface doesn't exhibits a sharpe change in $L=300^{\circ}$C when power prices are too low. In this case the surface becomes dependent on gas prices for boiler temperatures below the operating level. This dependence vanishes away for higher values of $S_e$.  

\begin{figure}
	\centering
	\begin{subfigure}[b]{0.55\textwidth}
		\includegraphics[width=\textwidth]{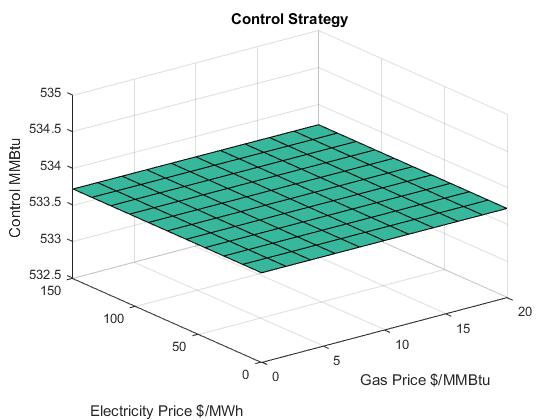}
		\caption{L=20}
		\label{F7a}
	\end{subfigure}%
	~ 
	\begin{subfigure}[b]{0.55\textwidth}
		\includegraphics[width=\textwidth]{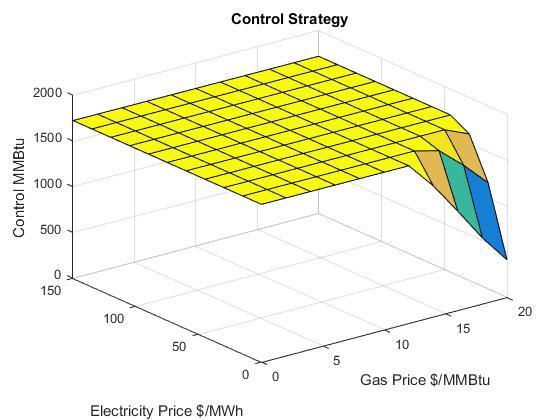}
		\caption{L=300}
		\label{F7b}
	\end{subfigure}
	
	\begin{subfigure}[b]{0.55\textwidth}
		\includegraphics[width=\textwidth]{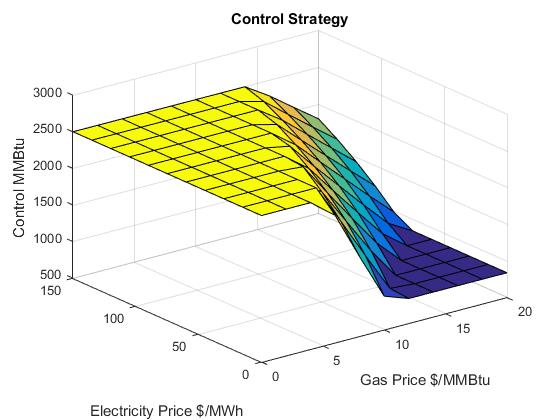}
		\caption{L=420}
		\label{F7c}
	\end{subfigure}%
	~ 
	\begin{subfigure}[b]{0.55\textwidth}
		\includegraphics[width=\textwidth]{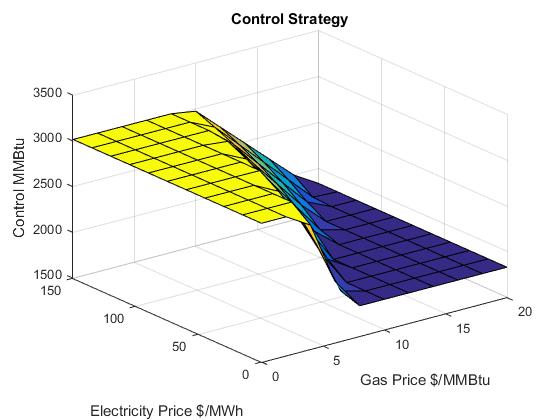}
		\caption{L=600}
		\label{F7d}
	\end{subfigure}
	\caption{Operating strategy for a single regime model w.r.t. electricity and gas spot prices for different boiler temperatures}\label{Fig7}
\end{figure}

The operating strategies w.r.t. electricity and gas spot prices for different boiler temperatures are depicted in Figure \ref{Fig7}. For very low boiler temperatures the control strategy surface is flat. In other words, it doesn't change depending on underlying spot prices. Starting from the case $L=300^{\circ}$C we notice a small region of dependence on power and gas spot prices. For higher temperature levels we can see that dependence more clearly (Figures \ref{F7c} and \ref{F7d}). 

We can differentiate 3 different regions here: $c_{\min}$ (dark blue), $c_{\max}$ (yellow), and $c_{min}\leq c \leq c_{max}$ (in between).

\begin{figure}[h!]
	\centering
	\includegraphics[width=0.8\textwidth]{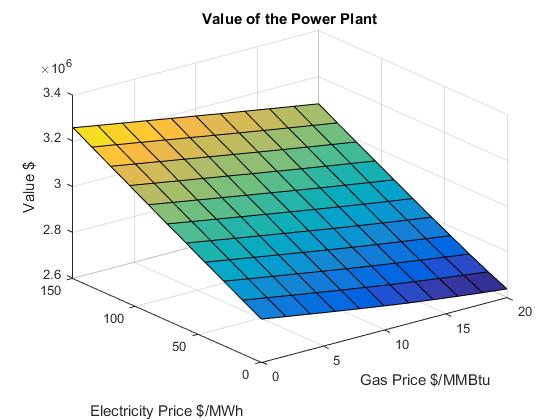}
	\caption{Value surface for the single regime model w.r.t. power and gas spot prices.}
	\label{Fig8}
\end{figure}

Figure \ref{Fig8} describes the power generator value surface  w.r.t. electricity and gas spot prices when $L=320^{\circ}$C. The value surface exhibits increasing and decreasing dependence on power and natural gas spot prices, respectively. The overall shape of the surface is the same for other temperature levels as well.

\section{Regime-switching}
Let's assume the economy has two phases/regimes (0 and 1) and the switch has some finite probabilities: $\lambda^{0\to 1}dt$ and
 $\lambda^{1\to 0}dt$. We will model this regime switching by the following two state continuous-time Markov chain $m(t)$ presented by \cite{chen2010implications}:

\begin{equation} \label{F52}
dm(t)=(1-m(t-))dN^{0\to 1}-m(t-)dN^{1\to 0}
\end{equation}
where $t-$ is the time infinitesimally before $t$, $N^{0\to1}$ and $N^{1\to 0}$ are independent Poisson processes with intensities 
 $\lambda^{0\to 1}$ and
 $\lambda^{1\to 0}$ ,respectively. So, within a regime $l=m(t-)$ the underlying electricity and gas spot prices follow:
\begin{align} \label{F53}
S^e(t)&=\Lambda_e^l (t)+X^e(t)+Y^e(t), \\  
\intertext{where} \notag
dX^e(t)&=-\alpha_e^l X^e(t)dt+\sigma_e^l dB^e(t), \\  \notag
dY^e(t)&=-\alpha_e^l Y^e(t) dt+dL_e^l(t),
\end{align}
and
\begin{align} \label{F54}
S^g(t)&=\Lambda_g^l (t)+X^g(t)+Y^g(t), \\ 
\intertext{where}  \notag
dX^g(t)&=-\alpha_ g^lX^g(t)dt+\sigma_g^l dB^g(t), \\  \notag
dY^g(t)&=-\alpha_g^l Y^g(t) dt+dL_g^l(t),
\end{align}
where $L_e^l$ and $L_g^L$ are  L\'{e}vy processes with intensity measures $\nu_e^l$ and $\nu_g^l$, respectively. In our compound Poisson case we have
\begin{equation*}
\nu_e^l(dz_e)=\lambda_e^l D_e^l(dz_e) \quad \text{and} \quad \nu_g^l(dz_g)=\lambda_g^l D_g^l(dz_g)
\end{equation*}

We can rewrite our power plant  optimisation objective \eqref{F22} for each regime $l$ as follows
\begin{equation} \label{F55}
V^l(S_e,S_g,L,t)=\max_{c^l(S_e,S_g,L,t)}\mathbb{E}[\int_t^T e^{-r(\tau-t)}(H(L)S_e-S_gc)d\tau|m(t)=l],
\end{equation}
subject to
\begin{equation*} 
c_{min}(L)\leq c^l \leq c_{max}(L).
\end{equation*}
The application of previously used techniques, such as Bellman's Principle of Optimality and It\^{o}'s Lemma will give us the following coupled HJB equations
\begin{multline} \label{F56}
V^l_{\tau}=\frac{1}{2}(\sigma_e^l)^2V^l_{S_e S_e}+
\frac{1}{2}(\sigma_g^l)^2V^l_{S_g S_g}+\rho\sigma_e^l\sigma_g^lV^l_{S_eS_g}+ 
\mu_e^l(S^e,T-\tau)V^l_{S_e}+\\
\mu_g^l(S^g,T-\tau) V^l_{S_g} -rV^l+\lambda^{l\to (1-l)}(V^{(1-l)}-V^l)\\
\int_{\mathbb{R}_+}(V^l_{S_e}(S_e+z_e,S_g,L,\tau)-V^l_{S_e}(S_e,S_g,L,\tau))\nu_e^l((dz_e) \\
+\int_{\mathbb{R}_+}(V^l_{S_g}(S_e,S_g+z_g,L,\tau)-V^l_{S_g}(S_e,S_g,L,\tau) )\nu_g^l(dz_g) \\
+\int_{\mathbb{R}_+^2}V^l_{S_e S_g}(S_e+z_e,S_g+z_g,L,\tau)F(U_1^l(z_e),U_2^l(z_g))dz_e dz_g \\
\max_c[H(L)S_e-S_gc+\eta(L,c)(\bar{L}(c)-L)V_L^l]
\end{multline}
with the terminal condition
\begin{equation*}
V^l(S_e,S_g,L,\tau=0)=0.
\end{equation*}
The boundary conditions at each regime will be the same as in Section \ref{Boundary}. 
\subsection{Numerical results}
We first need to define our input parameters for the regime switching model. We adopt the parameter values introduced in Section \ref{Secr0} for regime 0. Analogously, we make the following assumptions for regime 1 (see Table \ref{T2}).
\begin{table}
	\begin{center}
		\begin{tabular}{| l | l | l | l |}
			\hline
			Parameter & Value & Parameter & Value \\ \hline
			$\alpha_e$ & .6   & $\lambda_e$ & .2\\ \hline
			$\alpha_g$ & 1  &   $\lambda_g$ & .6\\ \hline
			$\sigma_e$ & .2  & $T$ & 200 hours \\
			\hline
			$\sigma_g$ & .3  & $\rho$ & .15  \\\hline
		\end{tabular}
	\end{center}
	\caption{Input parameters for the regime 1 of regime-switching model}
	\label{T2}
\end{table}
We assume that the risk-free interest rate is the same, $r=0.05$.  As the jump size distributions $D_e^1$ and $D_g^1$ we again take inverse Gaussian distributions with parameters $(.30,.46)$ and $(.42,.28)$, respectively. 

The seasonality terms $\Lambda_e^1$ and $\Lambda_g^1$ will be assumed to be
\begin{align*}
\Lambda_e^1(t)&=5 \sin((2\pi t-15.4\pi)/24)+10, \\
\Lambda_g^1(t)&=0.3 \cos(2\pi (t-18\pi)/24)+1.4.
\end{align*}
 Figures \ref{F9} and \ref{F10} compare operating strategies for the regimes 0 and 1, respectively w.r.t. power prices and boiler temperature for different gas spot prices. We can see that under both of these regimes the optimal strategy for $S_g \to 0$ is to burn maximum amount of gas all the time. However, for all other non-zero natural gas prices the region of maximum fuel consumption is larger for the regime 0 than for the regime 1. When gas prices are too high the plant no longer consumes fuel when electricity prices are low and $L<300^{\circ}$C. In other words, the plant manager becomes more pessimistic (or risk-averse) under the regime-switching model.
\begin{figure}
	\centering
	\begin{subfigure}[b]{0.55\textwidth}
		\includegraphics[width=\textwidth]{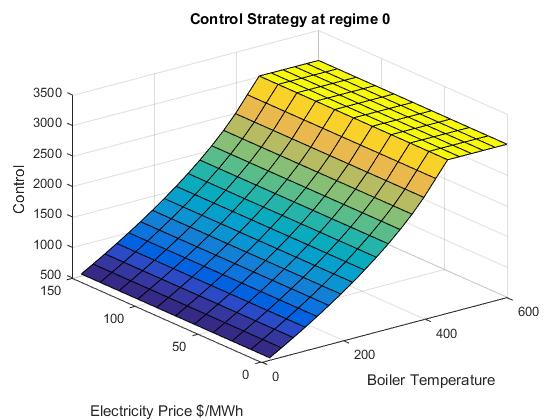}
		\caption{$S_g=0$}
		\label{F9a}
	\end{subfigure}%
	~ 
	\begin{subfigure}[b]{0.55\textwidth}
		\includegraphics[width=\textwidth]{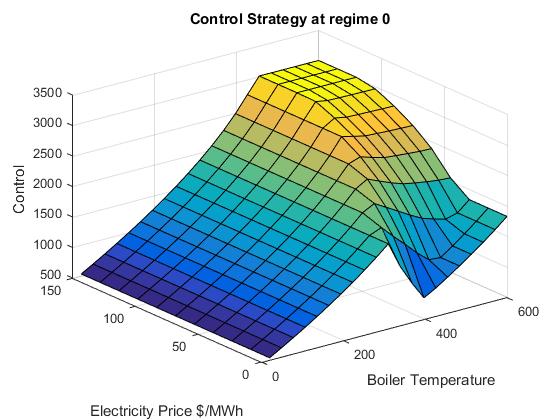}
		\caption{$S_g=10$}
		\label{F9b}
	\end{subfigure}
	
	\begin{subfigure}[b]{0.55\textwidth}
		\includegraphics[width=\textwidth]{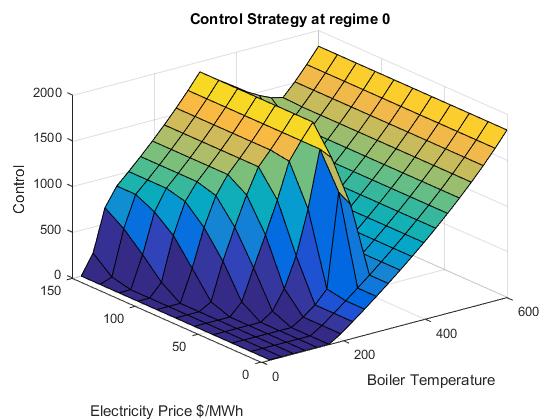}
		\caption{$S_g=14$}
		\label{F9c}
	\end{subfigure}%
	~ 
	\begin{subfigure}[b]{0.55\textwidth}
		\includegraphics[width=\textwidth]{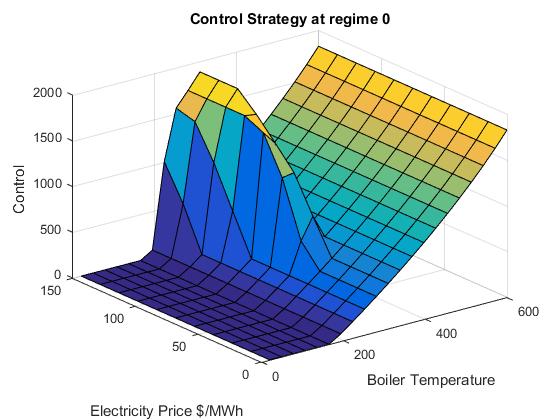}
		\caption{$S_g=20$}
		\label{F9d}
	\end{subfigure}
	\caption{Operating strategy for the regime 0 of the regime-switching model w.r.t. electricity spot price and boiler temperature for different gas prices}\label{Fig9}
\end{figure}

\begin{figure}
	\centering
	\begin{subfigure}[b]{0.55\textwidth}
		\includegraphics[width=\textwidth]{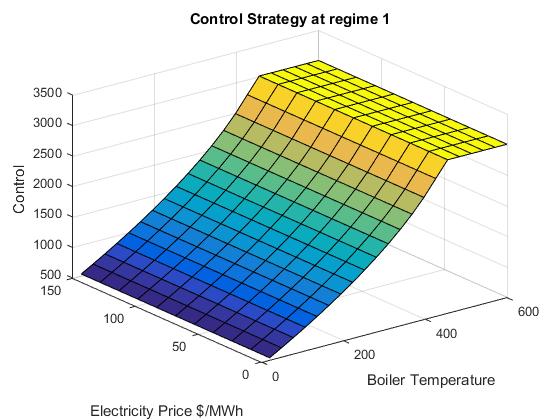}
		\caption{$S_g=0$}
		\label{F10a}
	\end{subfigure}%
	~ 
	\begin{subfigure}[b]{0.55\textwidth}
		\includegraphics[width=\textwidth]{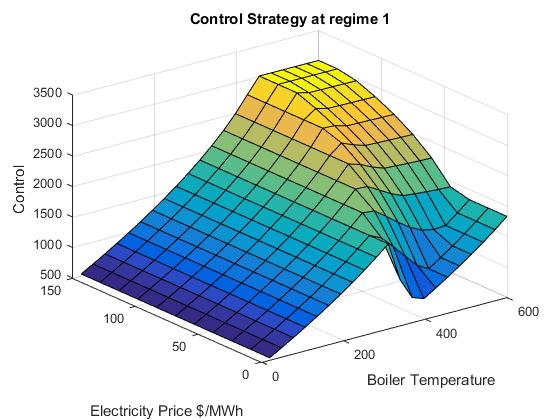}
		\caption{$S_g=10$}
		\label{F10b}
	\end{subfigure}
	
	\begin{subfigure}[b]{0.55\textwidth}
		\includegraphics[width=\textwidth]{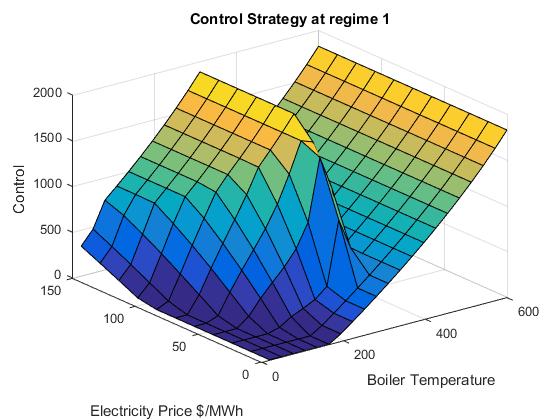}
		\caption{$S_g=14$}
		\label{F10c}
	\end{subfigure}%
	~ 
	\begin{subfigure}[b]{0.55\textwidth}
		\includegraphics[width=\textwidth]{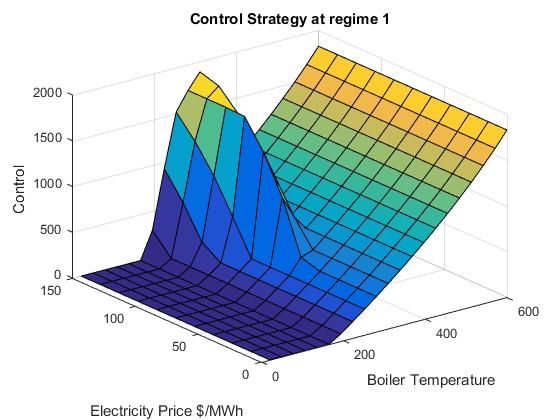}
		\caption{$S_g=20$}
		\label{F10d}
	\end{subfigure}
	\caption{Operating strategy for the regime 1 of the regime-switching model w.r.t. electricity spot price and boiler temperature for different gas prices}\label{Fig10}
\end{figure}

\begin{figure}
	\centering
	\begin{subfigure}[b]{0.55\textwidth}
		\includegraphics[width=\textwidth]{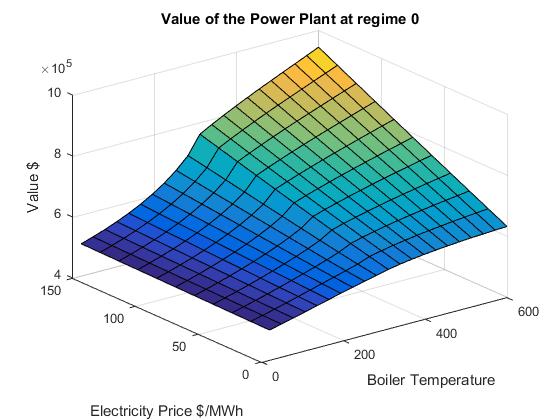}
		\caption{Regime 0}
		\label{F11a}
	\end{subfigure}%
	~ 
	\begin{subfigure}[b]{0.55\textwidth}
		\includegraphics[width=\textwidth]{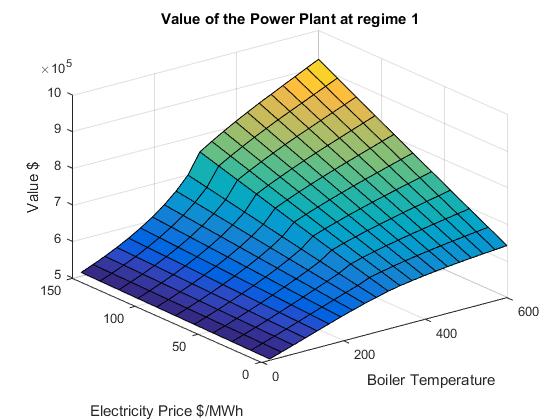}
		\caption{Regime 1}
		\label{F11b}
	\end{subfigure}
	\caption{Value surfaces for the regime-switching model w.r.t. power prices and boiler temperature for fixed gas price.}\label{Fig11}
\end{figure}

Value surfaces for the regime-switching model w.r.t. power prices and boiler temperature for the fixed gas price are depicted in Figure \ref{Fig11}. The surfaces look similar in their shapes but in regime 1 we have a downside shift which reflects an unfavorable economic situation.

\begin{figure}
	\centering
	\begin{subfigure}{0.32\textwidth}
		\centering
		\includegraphics[width=\textwidth]{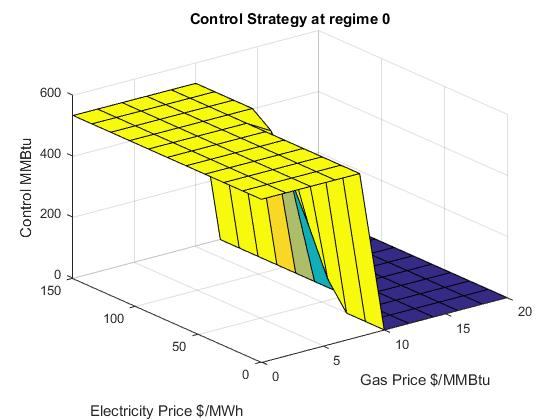}
		\caption{Regime 0, L=20}
		\label{F12a}
	\end{subfigure}%
	~ 
	\begin{subfigure}{0.32\textwidth}
		\centering
		\includegraphics[width=\textwidth]{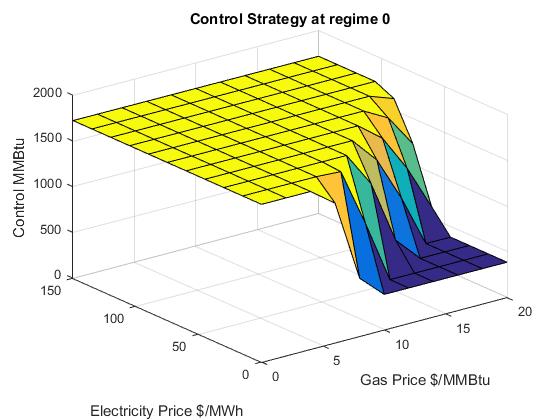}
		\caption{Regime 0, L=320}
		\label{F12b}
	\end{subfigure}
	\begin{subfigure}{0.32\textwidth}
		\centering
		\includegraphics[width=\textwidth]{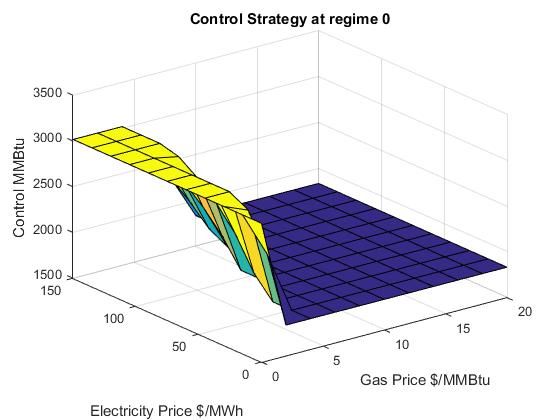}
		\caption{Regime 0, L=600}
		\label{F12c}
	\end{subfigure}%
	
	\begin{subfigure}{0.32\textwidth}
		\centering
		\includegraphics[width=\textwidth]{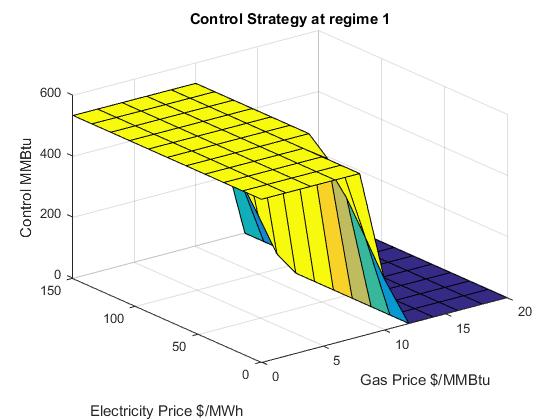}
		\caption{Regime 1, L=20}
		\label{F12d}
	\end{subfigure}%
	~ 
	\begin{subfigure}{0.32\textwidth}
		\centering
		\includegraphics[width=\textwidth]{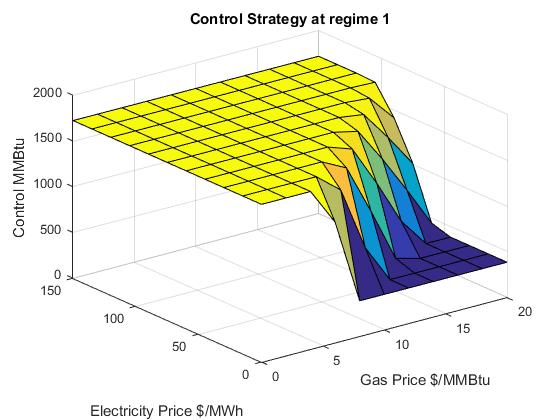}
		\caption{Regime 1, L=320}
		\label{F12e}
	\end{subfigure}
	\begin{subfigure}{0.32\textwidth}
		\centering
		\includegraphics[width=\textwidth]{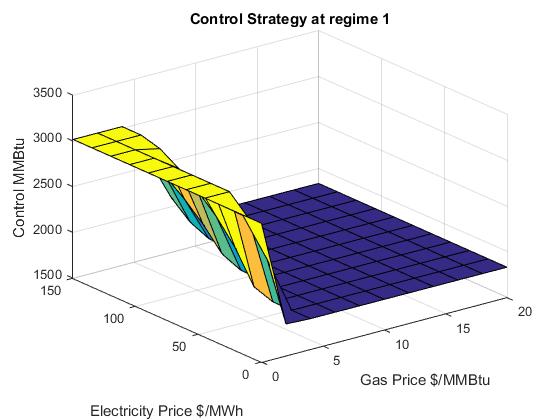}
		\caption{Regime 1, L=600}
		\label{F12f}
	\end{subfigure}%
	\caption{Operating strategy for regime-switching model w.r.t. electricity and gas spot prices for different boiler temperatures}\label{Fig12}
\end{figure}

Control strategies for regime-switching model w.r.t. power and gas spot prices for different boiler temperatures are presented in Figure \ref{Fig12}. When the temperature is above the operating level $L=300^{\circ}$C, optimal operation surfaces are more risk-averse in regime 1. As the plant doesn't produce electricity below 
$300^{\circ}$C, the change in power spot prices does not affect our decision making. Hence, in regime 1 the optimal control surface for $L=20^{\circ}$C is more optimistic than in regime 0 due to fall in fas spot prices (see Figures \ref{F12a} and \ref{F12d}). Moreover, in contrast to Figure \ref{Fig7}, the surface isn't flat when the temperature is close to its minimum.

\begin{figure}
	\centering
	\begin{subfigure}{0.32\textwidth}
		\centering
		\includegraphics[width=\textwidth]{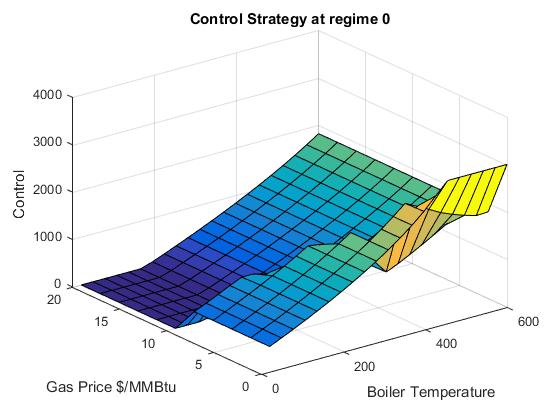}
		\caption{Regime 0, $S_e=0$}
		\label{F13a}
	\end{subfigure}%
	~ 
	\begin{subfigure}{0.32\textwidth}
		\centering
		\includegraphics[width=\textwidth]{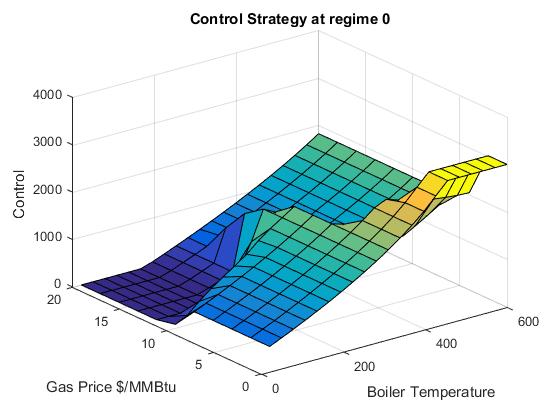}
		\caption{Regime 0, $S_e=60$}
		\label{F13b}
	\end{subfigure}
	\begin{subfigure}{0.32\textwidth}
		\centering
		\includegraphics[width=\textwidth]{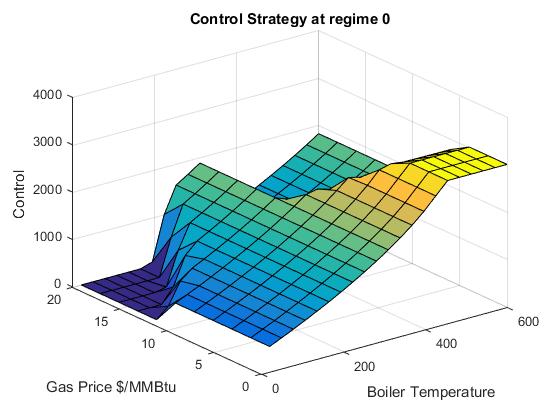}
		\caption{Regime 0, $S_e=150$}
		\label{F13c}
	\end{subfigure}%
	
	\begin{subfigure}{0.32\textwidth}
		\centering
		\includegraphics[width=\textwidth]{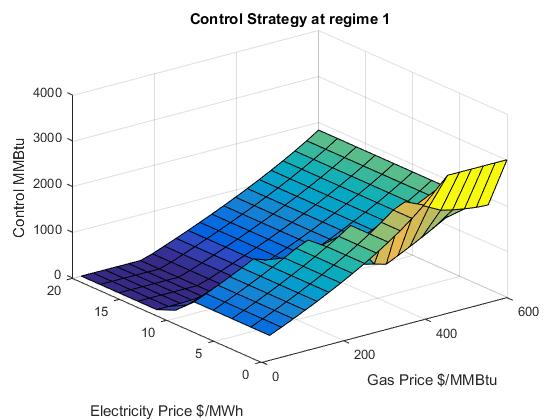}
		\caption{Regime 1, $S_e=0$}
		\label{F13d}
	\end{subfigure}%
	~ 
	\begin{subfigure}{0.32\textwidth}
		\centering
		\includegraphics[width=\textwidth]{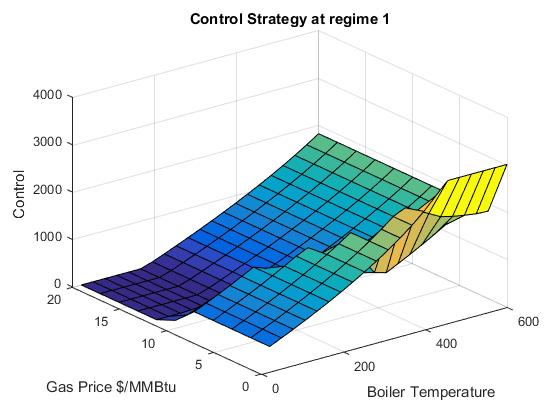}
		\caption{Regime 1, $S_e=60$}
		\label{F13e}
	\end{subfigure}
	\begin{subfigure}{0.32\textwidth}
		\centering
		\includegraphics[width=\textwidth]{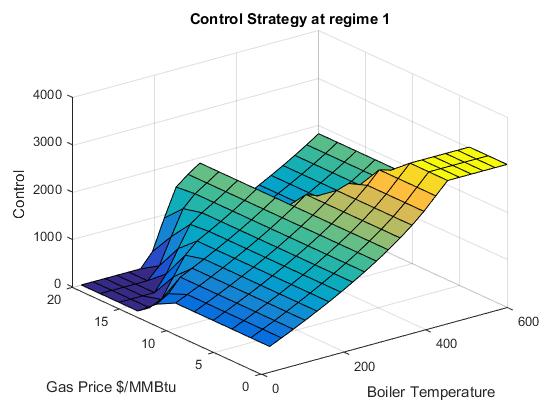}
		\caption{Regime 1, $S_e=150$}
		\label{F13f}
	\end{subfigure}%
	\caption{Operating strategy for regime-switching model w.r.t. gas spot prices and boiler temperature for different power prices}\label{Fig13}
\end{figure}

In Figure \ref{Fig13} we can observe operating strategy for regime-switching model w.r.t. gas spot prices and boiler temperature for different electricity spot prices.
In general, these figures show less appealing results than in single regime case (see Figure \ref{Fig6}). When the power prices are extremely low or high, the surfaces in regimes 0 and 1 look almost the same. However, for moderate electricity prices we have a larger region of minimal fuel consumption in regime 1.

\begin{figure}
	\centering
	\begin{subfigure}[b]{0.55\textwidth}
		\includegraphics[width=\textwidth]{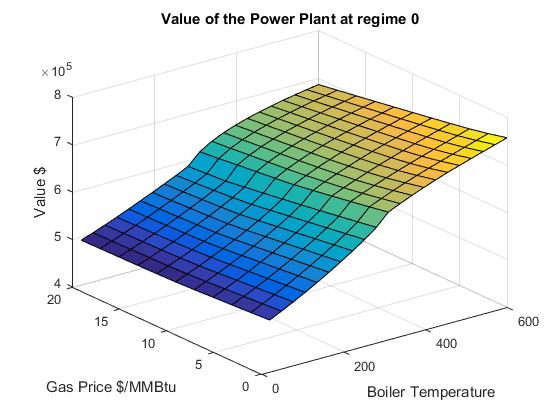}
		\caption{Regime 0}
		\label{F14a}
	\end{subfigure}%
	~ 
	\begin{subfigure}[b]{0.55\textwidth}
		\includegraphics[width=\textwidth]{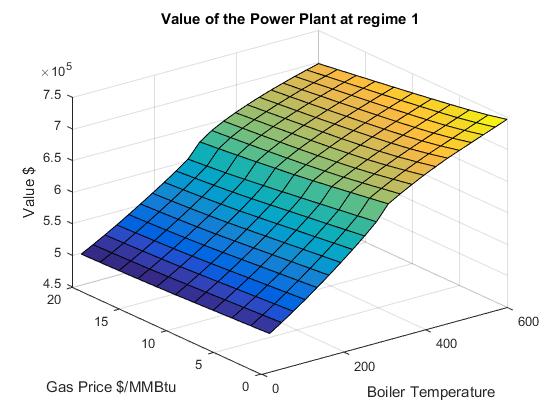}
		\caption{Regime 1}
		\label{F14b}
	\end{subfigure}
	\caption{Value surfaces for the regime-switching model w.r.t. gas spot prices and boiler temperature for fixed electricity price.}\label{Fig14}
\end{figure}

\begin{figure}
	\centering
	\begin{subfigure}[b]{0.55\textwidth}
		\includegraphics[width=\textwidth]{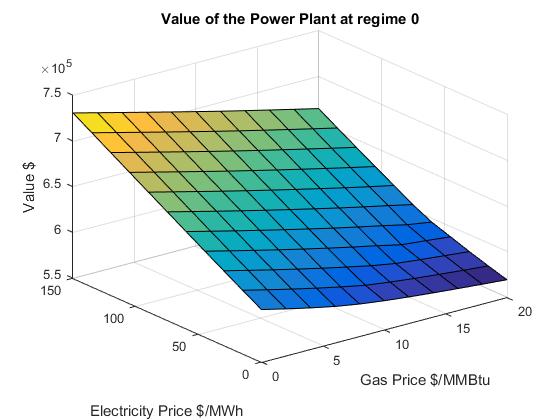}
		\caption{Regime 0}
		\label{F15a}
	\end{subfigure}%
	~ 
	\begin{subfigure}[b]{0.55\textwidth}
		\includegraphics[width=\textwidth]{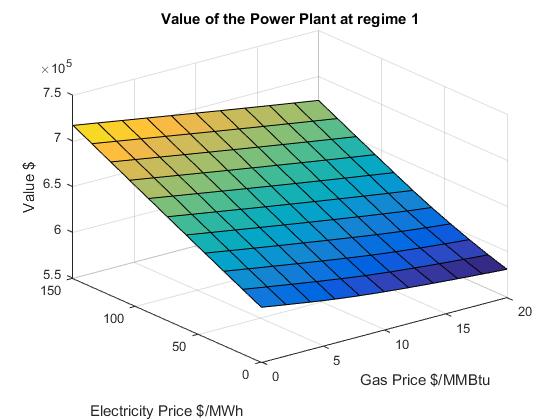}
		\caption{Regime 1}
		\label{F15b}
	\end{subfigure}
	\caption{Value surfaces for the regime-switching model w.r.t. electricity and gas spot prices for fixed boiler temperature.}\label{Fig15}
\end{figure}

Value surfaces for the regime-switching case w.r.t. gas spot prices and boiler temperature for fixed electricity price are provided in Figure \ref{Fig14}, while the corresponding surfaces w.r.t. electricity and gas spot prices are given in Figure \ref{Fig15}. We can infer from these figures that the value surfaces are shifted down in regime 1 due to economic disruptions. 

\section{Conclusion and future research plans}
In this research we analysed a stochastic control problem for gas-fired power generators while taking into account operating characteristics. The optimisation method introduced by \cite{thompson2004valuation} was elaborated by incorporating electricity and natural gas spot prices that exhibit regime-switching and interdependent spikes. The dependence of the price jumps has been modeled by means of skewed L\'{e}vy copulas.  We combined different numerical techniques to solve the resulting coupled non-linear HJB equations. We also used a minmod slope limiter to eliminate numerical oscillations from the optimal solution. The numerical approach gives us both the value of the power plant and optimal control strategy depending on the electricity and gas price, boiler temperature and time.  The numerical approach was implemented in MATLAB for a single regime and regime-switching cases. We investigated the differences in those models by observing the surfaces of optimal operation and value surfaces.

As the next step, we plan to extend our  numerical examples to the cases when underlying power and natural gas spot prices follow interdependent generalised hyperbolic (see \cite{eberlein2001application} and \cite{eberlein2002generalized}) and variance-gamma (see \cite{madan1998variance} or \cite{brody2012general}) L\'{e}vy models.The application of this optimisation approach into more complicated power generating assets could also be investigated.
\FloatBarrier
\bibliographystyle{plainnat}
\bibliography{energy_gasp}

%
%
%
%
%
\end{document}